\input harvmac
\noblackbox

%\KiritsisYV
\lref\KiritsisYV{
E.~Kiritsis and C.~Kounnas,
%``Curved four-dimensional space-times as infrared regulator in superstring theories,''
Nucl.\ Phys.\ Proc.\ Suppl.\  {\bf 41}, 331 (1995)
[arXiv:hep-th/9410212].
%%CITATION = HEP-TH 9410212;%%
}

%\KiritsisTA
\lref\KiritsisTA{
E.~Kiritsis and C.~Kounnas,
%``Infrared regularization of superstring theory and the one loop calculation of coupling constants,''
Nucl.\ Phys.\ B {\bf 442}, 472 (1995)
[arXiv:hep-th/9501020].
%%CITATION = HEP-TH 9501020;%%
}

%\BanksZY
\lref\BanksZY{
T.~Banks,
``On isolated vacua and background independence,''
arXiv:hep-th/0011255.
%%CITATION = HEP-TH 0011255;%%
}

%\FischlerCI
\lref\FischlerCI{
W.~Fischler and L.~Susskind,
``Dilaton Tadpoles, String Condensates And Scale Invariance,''
Phys.\ Lett.\ B {\bf 171}, 383 (1986).
%%CITATION = PHLTA,B171,383;%%
}

%\FischlerTB
\lref\FischlerTB{
W.~Fischler and L.~Susskind,
``Dilaton Tadpoles, String Condensates And Scale Invariance.
2,''
Phys.\ Lett.\ B {\bf 173}, 262 (1986).
%%CITATION = PHLTA,B173,262;%%
}

\lref\sjrey{S.~R.~Das and S.~J.~Rey,
``Dilaton Condensates And Loop Effects In Open And Closed Bosonic
Strings,''
Phys.\ Lett.\ B {\bf 186}, 328 (1987);
%%%CITATION = PHLTA,B186,328;
S.~J.~Rey, ``Off-Shell Prescription For Dilaton Tadpoles,''
Phys.\ Lett.\ B {\bf 203}, 393 (1988);
%%CITATION = PHLTA,B203,393;
S.~J.~Rey, ``Unified View Of BRST Anomaly And Its Cancellation In String
Amplitudes,''
Nucl.\ Phys.\ B {\bf 316}, 197 (1989).
%%CITATION = NUPHA,B316,197;
}

%\CallanWZ
\lref\CallanWZ{
C.~G.~Callan, C.~Lovelace, C.~R.~Nappi and S.~A.~Yost,
``Loop Corrections To Superstring Equations Of Motion,''
Nucl.\ Phys.\ B {\bf 308}, 221 (1988).
%%CITATION = NUPHA,B308,221;%%
}

%\BalasubramanianRE
\lref\BalasubramanianRE{
V.~Balasubramanian and P.~Kraus,
``A stress tensor for anti-de Sitter gravity,''
Commun.\ Math.\ Phys.\  {\bf 208}, 413 (1999)
[arXiv:hep-th/9902121].
%%CITATION = HEP-TH 9902121;%%
}

\lref\scottren{S. Thomas, in progress.}

%\MoffatCE
\lref\MoffatCE{
J.~W.~Moffat,
%``The cosmological constant problem and nonlocal quantum gravity,''
arXiv:hep-th/0207198.
%%CITATION = HEP-TH 0207198;%%
}
%\OoguriRJ
\lref\OoguriRJ{
H.~Ooguri and N.~Sakai,
``String Multiloop Corrections To Equations Of Motion,''
Nucl.\ Phys.\ B {\bf 312}, 435 (1989).
%%CITATION = NUPHA,B312,435;%%
}

%\CallanWZ
\lref\CallanWZ{
C.~G.~Callan, C.~Lovelace, C.~R.~Nappi and S.~A.~Yost,
``Loop Corrections To Superstring Equations Of Motion,''
Nucl.\ Phys.\ B {\bf 308}, 221 (1988).
%%CITATION = NUPHA,B308,221;%%
}

%\OoguriPK
\lref\OoguriPK{
H.~Ooguri and N.~Sakai,
``String Loop Corrections From Fusion Of Handles And Vertex
Operators,''
Phys.\ Lett.\ B {\bf 197}, 109 (1987).
%%CITATION = PHLTA,B197,109;%%
}

%\AdamsJB
\lref\AdamsJB{
A.~Adams and E.~Silverstein,
``Closed string tachyons, AdS/CFT, and large N QCD,''
Phys.\ Rev.\ D {\bf 64}, 086001 (2001)
[arXiv:hep-th/0103220].
%%CITATION = HEP-TH 0103220;%%
}

%\AharonyPA
\lref\AharonyPA{
O.~Aharony, M.~Berkooz and E.~Silverstein,
``Multiple-trace operators and non-local string theories,''
JHEP {\bf 0108}, 006 (2001)
[arXiv:hep-th/0105309].
%%CITATION = HEP-TH 0105309;%%
}

%\AharonyDP
\lref\AharonyDP{
O.~Aharony, M.~Berkooz and E.~Silverstein,
``Non-local string theories on $AdS_3 \times S^3$
and stable  non-supersymmetric backgrounds,''
Phys.\ Rev.\ D {\bf 65}, 106007 (2002)
[arXiv:hep-th/0112178].
%%CITATION = HEP-TH 0112178;%%
}

%\KachruYS
\lref\KachruYS{
S.~Kachru and E.~Silverstein,
``4d conformal theories and strings on orbifolds,''
Phys.\ Rev.\ Lett.\  {\bf 80}, 4855 (1998)
[arXiv:hep-th/9802183].
%%CITATION = HEP-TH 9802183;%%
}

%\LawrenceJA
\lref\LawrenceJA{
A.~E.~Lawrence, N.~Nekrasov and C.~Vafa,
``On conformal field theories in four dimensions,''
Nucl.\ Phys.\ B {\bf 533}, 199 (1998)
[arXiv:hep-th/9803015].
%%CITATION = HEP-TH 9803015;%%
}

%\BershadskyMB
\lref\BershadskyMB{
M.~Bershadsky, Z.~Kakushadze and C.~Vafa,
``String expansion as large N expansion of gauge theories,''
Nucl.\ Phys.\ B {\bf 523}, 59 (1998)
[arXiv:hep-th/9803076].
%%CITATION = HEP-TH 9803076;%%
}

%\KlebanovCH
\lref\KlebanovCH{
I.~R.~Klebanov and A.~A.~Tseytlin,
``A non-supersymmetric large N CFT from type 0 string
theory,''
JHEP {\bf 9903}, 015 (1999)
[arXiv:hep-th/9901101].
%%CITATION = HEP-TH 9901101;%%
}

%\TseytlinII
\lref\TseytlinII{
A.~A.~Tseytlin and K.~Zarembo,
``Effective potential in non-supersymmetric SU(N) x SU(N)
gauge theory  and interactions of type 0 D3-branes,''
Phys.\ Lett.\ B {\bf 457}, 77 (1999)
[arXiv:hep-th/9902095].
%%CITATION = HEP-TH 9902095;%%
}

\lref\fluxstab{}

%\BoussoXA
\lref\BoussoXA{
R.~Bousso and J.~Polchinski,
``Quantization of four-form fluxes and dynamical neutralization of the  cosmological constant,''
JHEP {\bf 0006}, 006 (2000)
[arXiv:hep-th/0004134].
%%CITATION = HEP-TH 0004134;%%
}

%\GiddingsYU
\lref\GiddingsYU{
S.~B.~Giddings, S.~Kachru and J.~Polchinski,
``Hierarchies from fluxes in string compactifications,''
arXiv:hep-th/0105097.
%%CITATION = HEP-TH 0105097;%%
}
%\KachruNS
\lref\KachruNS{
S.~Kachru, X.~Liu, M.~B.~Schulz and S.~P.~Trivedi,
``Supersymmetry changing bubbles in string theory,''
arXiv:hep-th/0205108.
%%CITATION = HEP-TH 0205108;%%
}

%\MaloneyRR
\lref\MaloneyRR{
A.~Maloney, E.~Silverstein and A.~Strominger,
``De Sitter space in noncritical string theory,''
arXiv:hep-th/0205316.
%%CITATION = HEP-TH 0205316;%%
}

%\FreyHF
\lref\FreyHF{
A.~R.~Frey and J.~Polchinski,
``N = 3 warped compactifications,''
Phys.\ Rev.\ D {\bf 65}, 126009 (2002)
[arXiv:hep-th/0201029].
%%CITATION = HEP-TH 0201029;%%
}

%\WittenUA
\lref\WittenUA{
E.~Witten,
``Multi-trace operators, boundary conditions, and AdS/CFT
correspondence,''
arXiv:hep-th/0112258.
%%CITATION = HEP-TH 0112258;%%
}

%\BerkoozUG
\lref\BerkoozUG{
M.~Berkooz, A.~Sever and A.~Shomer,
``Double-trace deformations, boundary conditions and spacetime
singularities,''
JHEP {\bf 0205}, 034 (2002)
[arXiv:hep-th/0112264].
%%CITATION = HEP-TH 0112264;%%
}

\lref\MStrassler{M. Strassler, to appear}

%\DvaliPE
\lref\DvaliPE{
G.~Dvali, G.~Gabadadze and M.~Shifman,
``Diluting cosmological constant in infinite volume extra
dimensions,''
arXiv:hep-th/0202174.
%%CITATION = HEP-TH 0202174;%%
}

%\PolchinskiRQ
\lref\PolchinskiRQ{
J.~Polchinski,
``String Theory. Vol. 1: An Introduction To The Bosonic String,''
{\it  Cambridge, UK: Univ. Pr. (1998) 402 p}.
}

%\FriedanAA
\lref\FriedanAA{
D.~Friedan,
``A tentative theory of large distance physics,''
arXiv:hep-th/0204131.
%%CITATION = HEP-TH 0204131;%%
}

\lref\IRgroup{N. Arkani-Hamed, S. Dimopoulos, G. Dvali, G.
Gabadadze, to appear.}

%\PolchinskiJQ
\lref\PolchinskiJQ{
J.~Polchinski,
``Factorization Of Bosonic String Amplitudes,''
Nucl.\ Phys.\ B {\bf 307}, 61 (1988).
%%CITATION = NUPHA,B307,61;%%
}

%\LaXK
\lref\LaXK{
H.~La and P.~Nelson,
``Effective Field Equations For Fermionic Strings,''
Nucl.\ Phys.\ B {\bf 332}, 83 (1990).
%%CITATION = NUPHA,B332,83;%%
}

%\AharonyCX
\lref\AharonyCX{
O.~Aharony, M.~Fabinger, G.~T.~Horowitz and E.~Silverstein,
``Clean time-dependent string backgrounds from bubble baths,''
JHEP {\bf 0207}, 007 (2002)
[arXiv:hep-th/0204158].
%%CITATION = HEP-TH 0204158;%%
}

%\ZwiebachIE
\lref\ZwiebachIE{
B.~Zwiebach,
``Closed string field theory: Quantum action and the B-V master equation,''
Nucl.\ Phys.\ B {\bf 390}, 33 (1993)
[arXiv:hep-th/9206084].
%%CITATION = HEP-TH 9206084;%%
}

%\OoguriRJ
\lref\OoguriRJ{
H.~Ooguri and N.~Sakai,
``String Multiloop Corrections To Equations Of Motion,''
Nucl.\ Phys.\ B {\bf 312}, 435 (1989).
%%CITATION = NUPHA,B312,435;%%
}

%\TseytlinVF
\lref\TseytlinVF{
A.~A.~Tseytlin,
``On 'Macroscopic String' Approximation In String Theory,''
Phys.\ Lett.\ B {\bf 251}, 530 (1990).
%%CITATION = PHLTA,B251,530;%%
}
%\DvaliFZ
\lref\DvaliFZ{
G.~Dvali, G.~Gabadadze and M.~Shifman,
``Diluting cosmological constant via large distance modification of  gravity,''
arXiv:hep-th/0208096.
%%CITATION = HEP-TH 0208096;%%
}

%\AharonyTT
\lref\AharonyTT{
O.~Aharony and T.~Banks,
``Note on the quantum mechanics of M theory,''
JHEP {\bf 9903}, 016 (1999)
[arXiv:hep-th/9812237].
%%CITATION = HEP-TH 9812237;%%
}

%\WittenMZ
\lref\WittenMZ{
E.~Witten,
``Strong Coupling Expansion Of Calabi-Yau Compactification,''
Nucl.\ Phys.\ B {\bf 471}, 135 (1996)
[arXiv:hep-th/9602070].
%%CITATION = HEP-TH 9602070;%%
}

\lref\higgsquestion{A. Adams, O. Aharony, J. McGreevy, E.
Silverstein, $\dots$, work in progress}

\def\subsubsec#1{\noindent{\it #1}}

\let\includefigures=\iftrue
\let\useblackboard=\iftrue
\newfam\black

%Figure Stuff
\includefigures
\message{If you do not have epsf.tex (to include figures),}
\message{change the option at the top of the tex file.}
\input epsf
\def\figin{\epsfcheck\figin}\def\figins{\epsfcheck\figins}
\def\epsfcheck{\ifx\epsfbox\UnDeFiNeD
\message{(NO epsf.tex, FIGURES WILL BE IGNORED)}
\gdef\figin##1{\vskip2in}\gdef\figins##1{\hskip.5in}% blank
space instead
\else\message{(FIGURES WILL BE INCLUDED)}%
\gdef\figin##1{##1}\gdef\figins##1{##1}\fi}
\def\DefWarn#1{}
\def\figinsert{\goodbreak\midinsert}
\def\ifig#1#2#3{\DefWarn#1\xdef#1{fig.~\the\figno}
\writedef{#1\leftbracket fig.\noexpand~\the\figno}%
\figinsert\figin{\centerline{#3}}\medskip\centerline{\vbox{
\baselineskip12pt\advance\hsize by -1truein
\noindent\footnotefont{\bf Fig.~\the\figno:} #2}}
\bigskip\endinsert\global\advance\figno by1}
%%%
\else
\def\ifig#1#2#3{\xdef#1{fig.~\the\figno}
\writedef{#1\leftbracket fig.\noexpand~\the\figno}%
%\figinsert\figin{\centerline{#3}}\medskip
%\centerline{\vbox{\baselineskip12pt
%\advance\hsize by -1truein\noindent
%\footnotefont{\bf Fig.~\the\figno:} #2}}
%\bigskip\endinsert
\global\advance\figno by1}
\fi
%

%%BLACKBOARD FONT STUFF
\useblackboard
\message{If you do not have msbm (blackboard bold) fonts,}
\message{change the option at the top of the tex file.}
\font\blackboard=msbm10 scaled \magstep1
\font\blackboards=msbm7
\font\blackboardss=msbm5
\textfont\black=\blackboard
\scriptfont\black=\blackboards
\scriptscriptfont\black=\blackboardss

\else

\fi
% *************************************
%\draft
%
\def\yboxit#1#2{\vbox{\hrule height #1 \hbox{\vrule width #1
\vbox{#2}\vrule width #1 }\hrule height #1 }}
\def\fillbox#1{\hbox to #1{\vbox to #1{\vfil}\hfil}}
\def\ybox{{\lower 1.3pt \yboxit{0.4pt}{\fillbox{8pt}}\hskip-
0.2pt}}
%
%
%%MATH MACROS
%Greek letters and their bars

%More bars

\def\comments#1{}

%AEL

%AEL

\def\II{\relax{I\kern-.10em I}}

\def\IZ{\relax\ifmmode\mathchoice
{\hbox{\cmss Z\kern-.4em Z}}{\hbox{\cmss Z\kern-.4em Z}}
{\lower.9pt\hbox{\cmsss Z\kern-.4em Z}}
{\lower1.2pt\hbox{\cmsss Z\kern-.4em Z}}
\else{\cmss Z\kern-.4emZ}\fi}
\def\IB{\relax{\rm I\kern-.18em B}}
\def\IC{{\relax\hbox{$\inbar\kern-.3em{\rm C}$}}}
\def\ID{\relax{\rm I\kern-.18em D}}
\def\IE{\relax{\rm I\kern-.18em E}}
\def\IF{\relax{\rm I\kern-.18em F}}
\def\IG{\relax\hbox{$\inbar\kern-.3em{\rm G}$}}
\def\IGa{\relax\hbox{${\rm I}\kern-.18em\Gamma$}}
\def\IH{\relax{\rm I\kern-.18em H}}
\def\II{\relax{\rm I\kern-.18em I}}
\def\IK{\relax{\rm I\kern-.18em K}}
\def\IP{\relax{\rm I\kern-.18em P}}
%\def\IX{\relax{\rm X\kern-.01em X}}
%this doesn't work

%

\def\inbar{\,\vrule height1.5ex width.4pt depth0pt}

\font\cmss=cmss10 
\def\IR{\relax{\rm I\kern-.18em R}}

%

 % for now

%

\def\lp10{\ell_p^{10}}
\def\lp11{\ell_p^{11}}
\def\R11{R_{11}}

\def\frac#1#2{{#1 \over #2}}

%%simeon's macros

%%\def\S{\Sigma} %%this is a bad idea
%%No it's not.  And stop commenting out
%%my macros without permission, john
%% -SH

\def\1dag{^{1\dagger}}
\def\2dag{^{2\dagger}}

\def\R#1#2#3{{{R_{#1}}^{#2}}_{#3}}

%%note the switch!!

%%ENGLISH MACROS
\def\ie{{\it i.e.}}
\def\cf{{\it c.f.}}

\hyphenation{Di-men-sion-al}
\def\eg{{\it e.g.}}

%%REFERENCING MACROS

%%

\Title{\vbox{\baselineskip12pt\hbox{hep-th/0209226}
\hbox{SU-ITP-02/36}
\hbox{SLAC-PUB-9504}
\hbox{PUTP-2048}}}
{
\vbox{
%\centerline{Decapitating Tadpoles in String Theory}}}
%\centerline{A Stringy Guillotine}
%\centerline{The Guillotine at the Boundary of Moduli Space}
%\centerline{This is Not a Tadpole}
%\centerline{A La Recherche du Tadpoles Perdu}
%\centerline{We Really Know Where Our Towels Are}
%\centerline{The Tadpole at the End of the Universe}
%\centerline{The Galactic Prophylactic:}
%\bigskip
%\centerline{Decapitating Tadpoles in QFT and String Theory}
%\centerline{Decapitating Tadpoles in Field and String Theory}
\centerline{Decapitating Tadpoles}
%\centerline{Stabilizing Non-SUSY Vacuua by}
%\bigskip
%\centerline{Decapitating Tadpoles}
%\centerline{Something from Nothing}
%\centerline{Beyond the Zero}
%\centerline{Elasticated Bloomers}
%\centerline{Radiatively Stable Asymptotically Flat String Vacuua}
%\bigskip
%\centerline{Without Supersymmetry}
%\centerline{}
}}
\bigskip
\bigskip
\centerline{Allan Adams$^{1}$, John McGreevy$^{1,2}$,
and Eva Silverstein$^{1}$}
\bigskip
\centerline{${}^{1}${\it Department of Physics and SLAC, Stanford
University, Stanford, CA 94305/94309}}
\bigskip
\centerline{${}^{2}${\it Department of Physics,
Princeton University, Princeton, NJ 08544}}
\bigskip
\bigskip
\noindent

%We argue that string theory can be modified consistently
%in the infrared to produce a unitary flat-space
%perturbative S-matrix even with broken spacetime supersymmetry.
%The modification is obtained by a perturbative bilocal
%deformation of
%the worldsheet action, of the sort found earlier
%in analyses of double-trace deformations of the AdS/CFT
%correspondence.
%In our flat space context, this modification is chosen
%so as to change the massless scalar and graviton propagators
%to decouple their zero modes, thereby removing
%divergences which would otherwise
%arise from their tadpoles.  The resulting S-matrix is
%parameterized by the VEVs of the scalars, which renders
%this class of string backgrounds rather unpredictive.  However,
%for generic values of these parameters, quantum effects
%produce masses for the nonzero modes of the scalars, lifting
%the fluctuating
%components of the moduli.

We propose that perturbative quantum field theory and string theory
can be consistently modified in the infrared to eliminate, in a
radiatively stable manner, tadpole instabilities that arise
after supersymmetry breaking.  This is achieved by deforming the
propagators of classically massless scalar fields
and the graviton so as to cancel
the contribution of their zero modes. In string theory, this
modification of propagators is accomplished
by perturbatively deforming the world-sheet action with bi-local
operators similar to those that arise in double-trace deformations
of AdS/CFT. This results in a perturbatively finite and unitary
S-matrix (in the case of string theory, this claim depends on
standard assumptions about unitarity in covariant string
diagrammatics). The S-matrix is parameterized by arbitrary scalar VEVs,
which exacerbates the vacuum degeneracy problem. However, for
generic values of these parameters, quantum effects produce masses
for the nonzero modes of the scalars, lifting the fluctuating
components of the moduli.
{\it Warning: in the case of string theory, the simple
prescription discussed in this paper fails to decouple BRST trivial
modes from the physical S-matrix.  A procedure aimed at correcting this
is under investigation.}

\bigskip
\Date{November, 2002}
%\draftmode

%-----------------------[ INTRODUCTION ]--------------------%
\newsec{Introduction and Summary}

Consider a flat space field theory or string theory with
one or more classically massless scalars.
After supersymmetry breaking, these scalars
(and the trace of the graviton) typically develop tadpoles at
generic points on the classical moduli space.
As a result, perturbation theory around generic points on the classical moduli
space does not produce a sensible S-matrix. This is because the
zero-momentum tadpole can attach itself to any diagram by the
massless propagator,
\eqn\divprop{
{1 \over k^2 } |_{k=0} =\infty,
}
rendering all amplitudes quantum mechanically divergent.
\ifig\ants{
In the presence of tadpoles, the flat space S matrix
does not exist due to divergences.
}{\epsfxsize3.0in\epsfbox{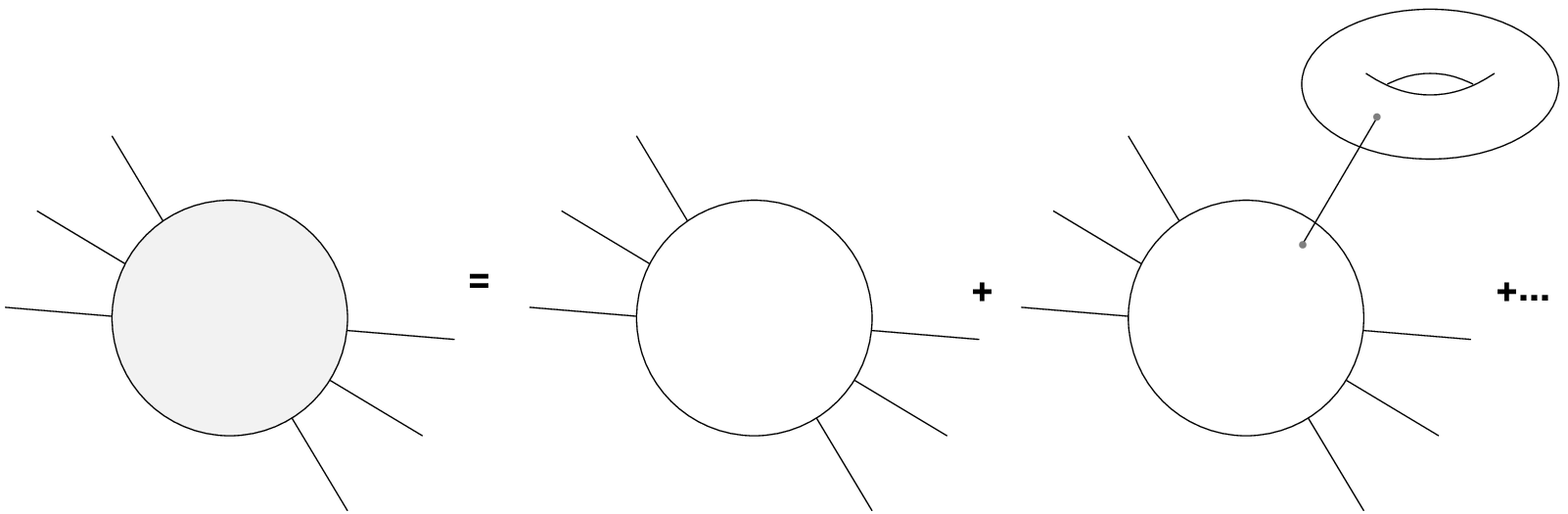}}
%%

%Similarly, quantum field theories and string theories experiencing
%dynamical supersymmetry breaking are subject to tadpole
%instabilities.

This IR divergence is usually  interpreted as a signal
that one must shift the massless field to an extremum of the
radiatively generated effective potential.
In string theory, this is
accomplished by adding the corresponding vertex operator to
the worldsheet action
\FischlerCI \FischlerTB (and \eg\ \sjrey\CallanWZ\OoguriRJ).  The
equations of motion satisfied by the
shifted field can be deduced cleanly from the
condition that BRST trivial modes decouple
in the string S-matrix \PolchinskiJQ\LaXK.\foot{In
the case where the scalar being shifted to its
extremum is the ubiquitous dilaton,
this often leads
to either
a trivial S-matrix, in the case that
the string coupling is driven to zero,
or a background which is not well described
by perturbation theory, in the case that the dilaton is
driven to strong coupling in some region of
spacetime.  In backgrounds of recent interest that fix
the dilaton at a nonzero value via flux stabilization or
nongeometrical monodromies, this problem may be avoided
(though so far in those cases
spacetime techniques have proven more practical than
worldsheet analysis).}

%

%

%Here, we would like to ask whether
%there is another way to produce
%a consistent (unitary) S-matrix replacing the nonexistent S-
%matrix in the
%original theory with unshifted tadpoles
We would like to suggest that there is another way to construct a
perturbatively consistent (\ie\ unitary) theory beginning with
this classical background.
Instead of shifting the massless fields, we will consider changing
their propagators. For example, for scalars, we will consider (an
IR and UV regulated version of)
\eqn\propchange{{-i\over{k^2+i\epsilon}}\to{-
i(1+F(k))\over{k^2+i\epsilon}} } where $F(k)$ is chosen to
preserve unitarity (and in string theory, worldsheet consistency
conditions) while satisfying $F(0)=-1$ in order to cancel the
contribution of the zero mode.\foot{Note that the tadpole only
sources the zero mode ($\int d^dx
\lambda_1\phi(x)=\lambda_1\phi_0$), as is clear diagrammatically
from the fact that energy momentum conservation forces the tadpole propagator
to $k=0$.}  This effectively changes the equations of motion
for the field whose tadpoles we are decapitating, so that any
point on the classical moduli space becomes a solution of
the deformed equations of motion.

This change is effected in string theory by the
perturbative application of the following non-local string theory
(NLST) \AharonyPA\AharonyDP\ deformation of the worldsheet action
(again to be regulated in the IR and UV in a manner to be explained in
detail in the body of the paper)
\eqn\basicdef{\delta S_{ws}=\int {d^dk\over
(2\pi)^d} { F(k)\over k^2+i\epsilon} \int V^{(k)}\int V^{(-k)}}
where $F(k)$ is chosen to have support only on-shell, on the
cone $k^2=0$. For simplicity, we will in fact take $F(k)$ to only have support at
$k=0$, though for scalars there may be other options,
and will define it as the limit of a smooth function. Here,
$\int V$ is an integrated vertex operator; the two factors of the
bilocal product can be inserted on the same Riemann surface or on
otherwise disconnected surfaces. Diagrammatically, each propagator
line is thus replaced by the right hand side of \propchange, so
here is the basic mechanism for removal (which we will refer to
as ``decapitation'') of tadpoles:
\ifig\deadants{
%Nonlocal method of birth control in string theory.
Cancellation of tadpole divergence via deformation of propagator.
The wedge denotes the contribution of the $F(k)$ term from
\basicdef\ in \propchange.
}{\epsfxsize3.0in\epsfbox{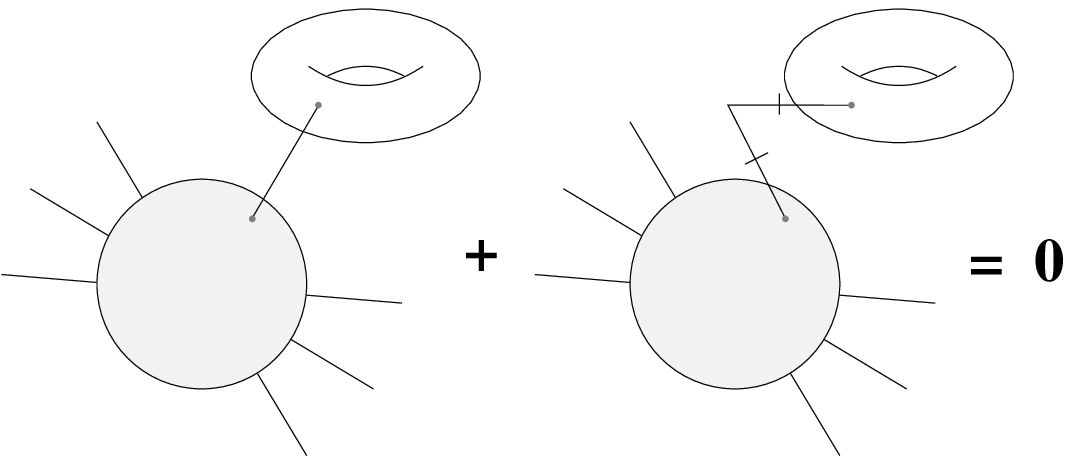}} % think: Clerks
%}{\epsfxsize3.0in\epsfbox{antdiagram.eps}}
In addition to decapitating dilaton and moduli scalar tadpoles,
we will decapitate the tadpole associated with the trace of the
zero momentum graviton in a similar way.

At the same time that we decapitate the tadpole, we remove the
zero modes of the massless scalars and graviton from the set of
external states we consider in the S matrix.
%in order
%to preserve the fact that every external state is associated
%with a pole in the S-matrix.
More generally we
will focus on the physical S matrix with generic incoming momenta
or with external states constructed from smooth wavepackets.  In
string theory, this is accomplished by rescaling the vertex
operators describing external states in our S-matrix by
\eqn\Vrescale{V^{(p)}\to\sqrt{1+F(p)}~V^{(p)}.}
%in order to preserve
%the fact that every external state is associated with a pole in
%the S-matrix.
We will choose $F(k)$ so that \basicdef\ does
not contribute to S-matrix elements
except via its cancellation of the massless tadpoles. This naively
ensures the perturbative unitarity of the resulting theory,
provided that the tadpole-free diagrams in the original theory
satisfy the cutting rules; this is manifest in simple field
theoretic examples and is thought to hold in superstring perturbation
theory.

\lref\StringyDeus{
Guardian of the Purity of String Theory, Private Communication
}

{\it
However, in string theory, simply removing the divergences is not enough to
ensure the perturbative unitarity of the resulting diagramatic
expansion, as pointed out in the context of this construction
by Joe Polchinski \StringyDeus.
%\foot{We are extremely grateful to Joe Polchinski
%\StringyDeus\ for bringing this subtlety of
%worldsheet perturbation theory to our attention, as well as for discussions
%on the issues it raises for the construction outlined in this paper.}.
In amplitudes with BRST trivial vertex operators, the undeformed
theory has a finite contribution from the tadpole which is
not cancelled by our deformation \basicdef\ as it stands.
Therefore the claims of consistency made in the remainder
of this paper on the string theory case based
on \basicdef\ alone are wrong.
An additional set of NLST deformations which cancel the BRST anomaly
as well as the divergences are
under investigation to see whether they lead to a fully
consistent theory.
While the procedure outlined in the remainder of this paper does not
result in a unitary string S-matrix for the string
theory case, it is worth noting that this problem
does not arise for the field theory case of our procedure.
}

As we will explain in detail in the bulk of the paper, this
effectively removes the {\it spacetime average} of the tadpole for
the field in a radiatively stable way, while retaining the
quantum-generated self energy for nonzero-momentum modes,
including mass renormalization lifting
moduli.  In simple examples (where the tadpole {\it is} constant
in spacetime) this leads to a nontrivial nonsupersymmetric
perturbative S-matrix in flat space. We will study this explicitly
for theories for which the tadpole is generated
perturbatively.\foot{We expect that similar results will hold in
situations with dynamical supersymmetry breaking.} The S matrix so
constructed agrees at tree level
with the classical S matrix of the undeformed
theory, but exists quantum mechanically (at least in perturbation
theory).  In this S-matrix the fluctuating (nonzero) modes of the
moduli are lifted, while the zero mode values (VEVs) of the moduli
constitute parameters (couplings) on which the S-matrix amplitudes
depend.

In quantum field theory, the perturbative S matrix we construct
this way is equivalent (for external states carrying generic
momenta or arranged into smooth wavepackets) to that which
one would
obtain from simply fine tuning away order by order the linear term
in the potential expanded about any value for the VEV of the
scalar field (or fine tuning away the cosmological constant in the
case of gravity). Such a prescription would not be radiatively
stable. Our prescription of a nonlocal shift in the propagator is
radiatively stable.  So, by enlarging the space of possible
backgrounds to include nonlocal deformations, one can realize in a
radiatively stable manner a system which would otherwise require
unnatural fine tuning. In perturbative string theory, one cannot
directly fine tune the spacetime effective action in any case, but
the decapitation prescription \basicdef\ can be
implemented directly and again provides the same effect in a
radiatively stable way.  It is also worth noting that it seems
likely that the full theory in the presence of $F(k)$,
including the possibility of
expanding around backgrounds other than flat space, is not
equivalent to that which one would obtain from fine tuning away the
tadpole.

\ifig\figCounterTerms{
A counterterm for the tadpole requires delicate order-by-order fine-tuning,
and depends critically on the UV cutoff.  By contrast,
decapitation automatically generates contributions cancelling the
tadpoles
to all orders once the tree level deformation has been
specified, and thus does not involve fine tuning.
}
{\epsfxsize3.0in\epsfbox{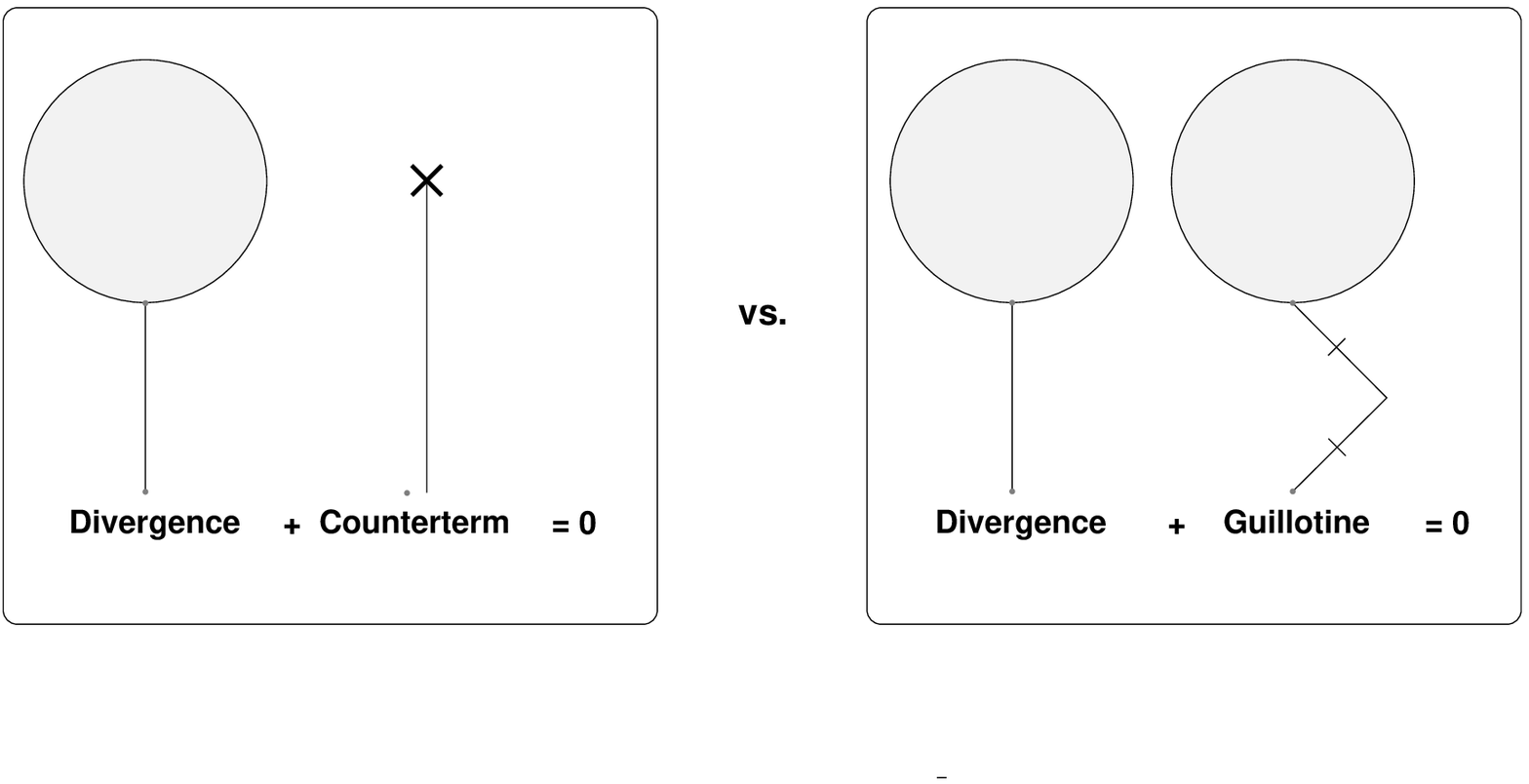}}

Even if we focus on the radiatively stable description in terms of
the modified tree-level propagator, we cannot regard this
prescription as a solution to the cosmological constant problem
per se since in the real world the tadpole is not constant in
spacetime.  Our prescription removing the zero mode does not
address the issue of phase transitions (variation in time)
and does not cancel the cosmological term in different localized
spatial domains (variation in space) \IRgroup. Indeed, one of the
appealing features of our construction is that the metric responds
normally to localized sources of stress-energy; it is only the
tadpoles due to the cosmological term which are removed by the
procedure.
It will be interesting to explore more
systematically the space of consistent IR modifications,
and to try to implement in string theory deformations with
a better chance of solving the real-world vacuum energy problem.

Finally, it should be mentioned that, although we will argue for
perturbative consistency (unitarity) of our S-matrix, we will
have nothing to say here about nonperturbative stability
and consistency.

Our argument may appear at odds with standard assumptions about
the unity and predictivity of string theory, which are supported
by some spectacular results of recent years.
Ordinary string/M theory has been unified significantly
by string dualities, and formulated nonperturbatively
in some backgrounds
by matrix theory and AdS/CFT.  However, these
beautiful results, while conceptually unifying the framework,
have not yet rendered the theory highly predictive.  Indeed,
the space of a priori possible string phenomenologies has grown tremendously
with the advent of nonperturbative gauge symmetries, D-branes,
and dual descriptions of large N gauge sectors; focusing
on elegant possibilities such as \WittenMZ\ may be well
motivated from phenomenological considerations and simplicity
but has not yet been seen as a prediction of the full theory,
which can apparently accommodate arbitrarily large gauge groups and matter
content.
In addition, the different backgrounds of the theory, while
mathematically arising from a unified framework, may not be
physically connected due to their very different UV and/or IR
behavior \AharonyTT\BanksZY.  In the context
of AdS/CFT the equivalence of quantum field theory and string
theory shows that string theory need not be more predictive
than field theory.  In the context of string compactification there is
growing evidence that many quantities in the low energy theory can
be effectively tuned by choosing the background
\BoussoXA\GiddingsYU\MaloneyRR.  The most urgent issue in evaluating
a potential new class of backgrounds of string theory is its
physical consistency.  The question of vacuum selection in the
full quantum theory is an issue that must certainly be addressed
but may well fall outside the scope of perturbation theory.
In any case, if our backgrounds can ultimately
be eliminated by some concrete physical consistency requirement going
beyond those we address in this paper, it
would serve as further evidence for the unity and predictivity of
string theory.
%%Long live Ralph Nader.

Regardless, our proposal, which will be checked in detail the bulk of this
paper, may seem outlandish on first sight.  Let us begin
therefore by sharing some of the motivations leading
to this idea, before embarking on a systematic analysis
of our prescription and its physical features.

%-----------------------[ Motivation from AdS/CFT ]--------------------%

\subsec{Motivation from AdS/CFT double-trace couplings}

Bilocal deformations of the general form of \basicdef, namely
\eqn\gendef{\delta S_{ws}\sim \sum_{I,J}c_{IJ} \int V^{(I)}\int
V^{(J)}} have been derived perturbatively on the string theory
side of AdS/CFT dual pairs perturbed by double trace deformations
\AharonyPA\AharonyDP. In some AdS/CFT examples
\KachruYS\LawrenceJA\BershadskyMB\KlebanovCH\WittenUA\BerkoozUG ,
running marginally-relevant double-trace couplings on the field
theory side are generated dynamically
\AdamsJB\TseytlinII\WittenUA\BerkoozUG\MStrassler\ and affect
some amplitudes in the theory at large $N$ \AharonyPA\AdamsJB.

On the field theory side, the space of couplings includes both
single-trace and arbitrary multitrace deformations.  These
couplings are all on the same footing in field theory (aside
from their effect on the structure of the 't Hooft expansion).
In specifying a field theory, one chooses a renormalization
group
trajectory accounting for the behavior of all the couplings.
Depending on how one organizes the perturbation expansion, this
may involve cancelling divergent amplitudes with counterterms.
The coefficients of these
counterterms are determined by appropriate renormalization
conditions.

Applying the dictionary of \AharonyPA, this suggests that one
should
enlarge the space of string backgrounds one considers to
include
those deformed from ordinary string theory by perturbations
of the form \gendef.  As in field theory, and in the case of
local deformations of string theory, appropriate
consistency
conditions will restrict this space of backgrounds to a
physical subspace.

Moreover, in the context of AdS/CFT, UV divergences requiring
counterterms
on the field theory side map to IR divergences on the string
theory side.  These IR divergences may therefore entail
a corresponding
renormalization prescription, including contributions of the
form \gendef\
required to cancel divergences, similarly to the way
counterterms
for double-trace couplings
cancel UV divergences on the field theory side
\BalasubramanianRE\scottren.

This idea is difficult to apply directly in the context of
AdS/CFT with dynamically generated double-trace interactions
in perturbation theory,
because of the usual difficulty involved in describing
the string theory side at large curvature.
In this paper, we will take this as motivation and
apply these ideas directly to flat space string theory,
studying the deformation of the form \basicdef\ and placing on
its coefficient $F(k)$ appropriate ``renormalization
conditons''
to ensure the finiteness and consistency of the resulting
S matrix.\foot{Another approach to flat space was adopted in
\AharonyDP, by taking a scaling limit of double-trace deformed
AdS/CFT to flat space; there one found
divergences from insertion of a bilocal product
of 0-momentum vertex operators, not smoothed by
an integral over $k$ as we have done in \basicdef.
In \AharonyCX, NLST deformations naturally
arose in describing the squeezed states obtained
from particle creation in an asymptotically
flat time-dependent background; again this
is different because our deformation \basicdef\
involves both positive and negative frequency modes and
does not constitute a squeezed state in the original
flat space background.}

%-----------------------[ Outline ]--------------------%

\subsec{Outline of the paper}

In section 2 we will present our prescription in detail and show
how it cancels tadpole divergences in a radiatively stable manner
and lifts the nonzero modes of the moduli.  In section 3 we will
address the question of other effects of the deformation, and show
that the deformation does not contribute for generic external
momenta (and therefore smooth wavepackets) to S-matrix elements
except via its cancellation of massless tadpoles. This in
particular ensures spacetime unitarity and Lorentz invariance of
the resulting S-matrix, given plausible assumptions about
superstring perturbation theory.  In section 4 we will assemble and discuss
some basic physical features of the construction, and discuss many
future directions.

%-----------------------[ Related Work ]--------------------%
\subsec{Related work}

The notion of modifying gravity in the IR and generalizing
renormalization to that context is an old idea which has
also been explored recently
in \DvaliPE\FriedanAA\DvaliFZ\IRgroup\MoffatCE.
The work \DvaliPE\DvaliFZ\ has pursued the possibility
of a consistent modification of gravity in the IR arising in
a brane configuration in a higher dimensional
bulk spacetime in the presence
of an Einstein term with large coefficient on the brane
worldvolume.
The work \IRgroup\ has
provided many insights into the requirements an IR
modification of gravity
must satisfy in order to be able to address the cosmological
constant problem including the effects of
phase transitions, while maintaining consistency with known
physics, and has proposed concrete examples and mechanisms
for satisfying
these requirements.
It would be interesting if an NLST prescription such as the
one we employ here to produce a consistent flat-space
nonsupersymmetric
S-matrix could provide a way to formulate a
consistent string-theoretic
embedding of the effective field theory
examples of \IRgroup.  The approach of
\FriedanAA\ is complementary to ours in a sense we will remark
on in the following.  Bilocal worldsheet terms appeared
in the work \TseytlinVF\ on generating effective field
theory from string theory, as well as in the more recent
context of the AdS/CFT double trace
deformations just reviewed.

%-----------------------[ THE PERSCRIPTION ]--------------------%

\newsec{The prescription, and cancellation of tadpole
divergences}

In this section we will lay out in detail the prescription
motivated and summarized
in the last section.

%-----------------------[ Tadpoles and Regulators ]--------------------%
\subsec{Tadpoles, Divergences, and Regulators}

%As reviewed in \PolchinskiRQ\ (chapter 9) one can parameterize the
%moduli space of string diagrams with ``plumbing fixtures''. In
%this description,

In string theory, similarly to field theory, the contribution of a
massless tadpole to an S-matrix element is by a factor of the zero
momentum propagator $G_2(k=0)$
%\propI\Tprop\
times the one-point function of the massless vertex operator at
zero momentum.  This multiplies the rest of the diagram given by
one insertion of the massless vertex operator at zero momentum
along with the insertions of vertex operators describing the
external states in the amplitude,
\eqn\factorampl{ {\cal
A}^h|_{Tadpole}\sim \langle \int V^{(k_1)}_1\dots\int
V^{(k_n)}_n\int V^{(0)}\rangle _{\Sigma_{\tilde h}} \times
G_2(k=0)\times \langle \int V^{(0)}\rangle _{\Sigma_{h-\tilde
h}}.}
This is represented diagramatically
%as a boundary of the moduli
%space of the $h$-loop diagram
as follows:
\ifig\ants{a) a generic $h$-loop amplitude; b) the contribution of
the one-loop tadpole to this amplitude is a product of the
$(h-1)$-loop amplitude, the one-loop tadpole, and a propagator.}
{\epsfxsize3.0in\epsfbox{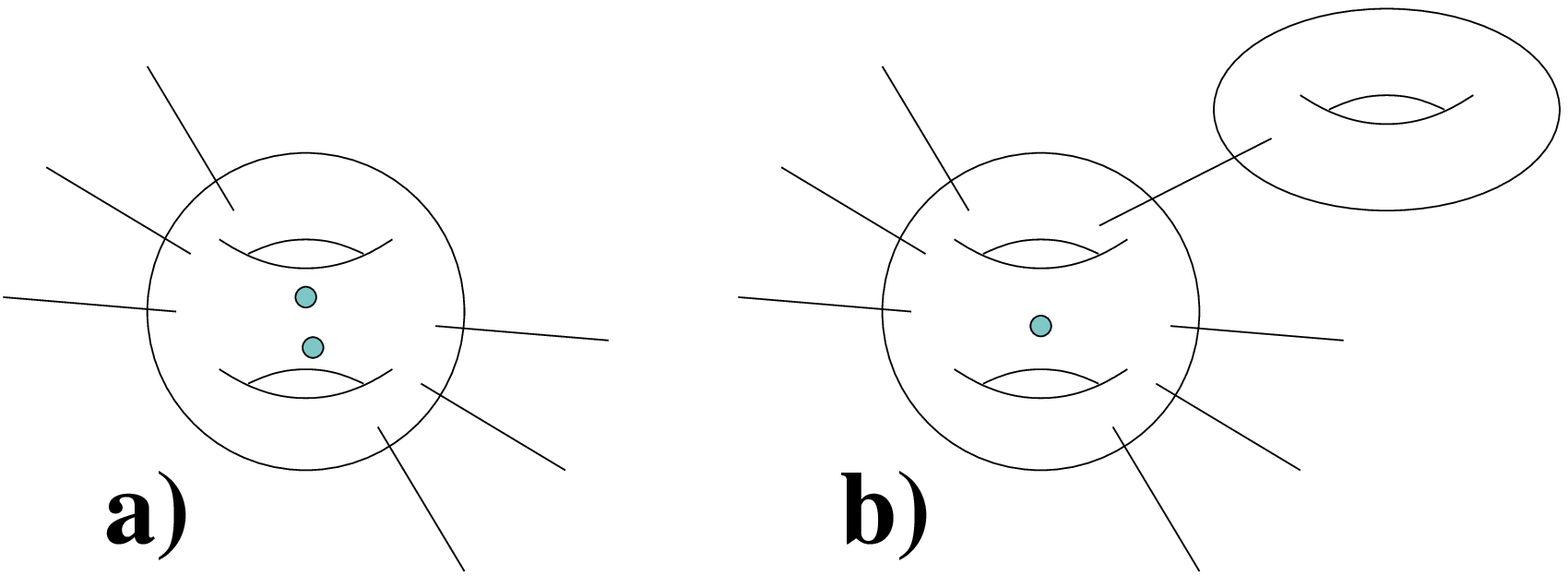}}
At least in the bosonic string, both the massless tadpole diagram
and the remaining contributions to the amplitude can be
represented as a collection of field theoretic diagrams
constructed from (an infinite number of) hermitian irreducible
vertices and propagators \ZwiebachIE\PolchinskiRQ.
In this decomposition, all spacetime IR divergences
arise from propagator contributions, not from the effective
vertices.
In the
superstring we expect a similar decomposition to hold, and we will
assume this, though to our knowledge this has not been proven.
This field theoretic decomposition will be important in
the following, particularly for our
analysis of unitarity in \S3\ (as in \PolchinskiRQ).

While the tadpole is finite in the absence of tachyons (thanks to
the soft UV properties of string loops) the on-shell massless
propagator is divergent, and requires regularization. We will
discuss two natural ways to do this in the case of scalar fields,
one of which generalizes to the graviton. We will work in
signature $(+,-,\dots,-)$, and denote by $d$ the
number of dimensions in
which the field whose tadpole we are decapitating propagates.

We begin by discussing classically massless scalar fields. In
field theory, a simple method of IR regulation, in situations
where it is consistent with gauge invariance, is the by-hand
introduction of a small mass $\mu$ to be taken to zero at the end
of each calculation
%
%computation\foot{As long as the field is not charged under any
%gauge symmetry, this is an unambiguous perscription.},
%
giving the regulated propagator,
\eqn\cuttransl{
{1\over k^2-\mu^2+i\epsilon}.
}

In string theory, infrared regularization is most directly
expressed in terms of a cutoff on the appropriate Schwinger
parameter arising in the propagator of the field theory
decomposition summarized above
(for a discussion of IR regulation in string theory, see e.g. 
\KiritsisYV\KiritsisTA).
 In particular, the
closed string
propagator is \eqn\propI{\lim_{T_c\to \infty, T_0\to 0}~ \int
\!\!\!\!\!\!\!\!\!\!\!
 \sum_{states}\int_{T_0}^{T_c} dT e^{-
T(L_0+\tilde L_0)} } In flat space, for a state corresponding to a
spacetime excitation with mass $m$ and momentum $k$ this gives
\eqn\Tprop{\eqalign{ G_2(k;T_c, T_0)&\sim \int_{T_0}^{T_c}dT
e^{T(k^2-m^2+i\epsilon)}\cr &={1\over
k^2-m^2+i\epsilon}\biggl(e^{T_c(k^2-m^2+i\epsilon)} -
e^{T_0(k^2-m^2+i\epsilon)}\biggr) }} Taking $T_c\to\infty$,
$T_0\to 0$ reproduces the usual pole ${1\over k^2-m^2+i\epsilon}$.
For finite (but large) $T_c$, as $k^2\to m^2$ we obtain an IR
regulated result \eqn\Gcutoff{ G_2(k^2\to m^2; T_c)\sim T_c } One
may define these momentum integrals in appropriate circumstances
by Euclidean continuation; in that case, $T_0$ represents a UV
cutoff which we may also employ. We can relate the two regulation
schemes near the IR limit $k\to 0$ by taking $T_c$ to be a
function of $k^2$ and $\mu^2$ given by the solution to
\eqn\cuttransl{ {1\over
k^2+i\epsilon}\biggl(e^{T_c(k^2+i\epsilon)} -
e^{T_0(k^2+i\epsilon)}\biggr) \equiv {1\over k^2-\mu^2+i\epsilon}.
}
%We will switch between regulation schemes rather freely.
We will mostly consider the hard (\ie\ $\mu$-independent) $T_c$
regulator, but will use the $\mu$ regulator in sufficiently simple
quantum field theory examples.

%-----------------------[ The Deformation ]--------------------%
\subsec{The Deformation}

We will consider our deformation both in perturbative
quantum field theory and string theory.
In the $\mu$ regularization scheme in quantum
field theory, we deform the propagator by
\eqn\basicdefmu{
{i F(k) \over k^2-\mu^2+i\epsilon}.
}
where $F(k)$ will be specified shortly.
One can also employ the Schwinger parameterization and
regularization in quantum field theory.

In string theory, in terms of the Schwinger
cutoff, we implement the following NLST deformation,
adding to the worldsheet action
\eqn\TNLST{ \delta
S_{ws}\propto \int d^dk {F(k)\over k^2+i\epsilon} (e^{T_c
(k^2+i\epsilon)} -e^{T_0 (k^2+i\epsilon)}) \int V^{(k)} \int
V^{(-k)} }
%
%adding to the worldsheet action \eqn\basicdefmu{\delta S_{ws}=\int
%{d^dk\over (2\pi)^d} { F(k)\over k^2-\mu^2+i\epsilon} \int
%V^{(k)}\int V^{(-k)},}
%
where $\int V$ are integrated vertex operators corresponding to
the massless particles whose tadpoles we wish to decapitate.

%In terms of the Schwinger
%cutoff, this NLST deformation reads \eqn\TNLST{ \delta
%S_{ws}\propto \int d^dk {F(k)\over k^2+i\epsilon} (e^{T_c
%(k^2+i\epsilon)} -e^{T_0 (k^2+i\epsilon)}) \int V^{(k)} \int
%V^{(-k)} }

As in \AharonyPA\AharonyDP, we treat this deformation
perturbatively.  This introduces an infinite array of
new diagrams in which the vertex operators in \TNLST\ attach to
Riemann surfaces in all possible combinations (including
diagrams in which the two members of the bi-local pair of vertex
operators sit on different, otherwise disconnected, Riemann
surfaces).
\ifig\ants{
The bi-local deformation can connect two Riemann surfaces
or attach to a single Riemann surface.
}
{\epsfxsize3.0in\epsfbox{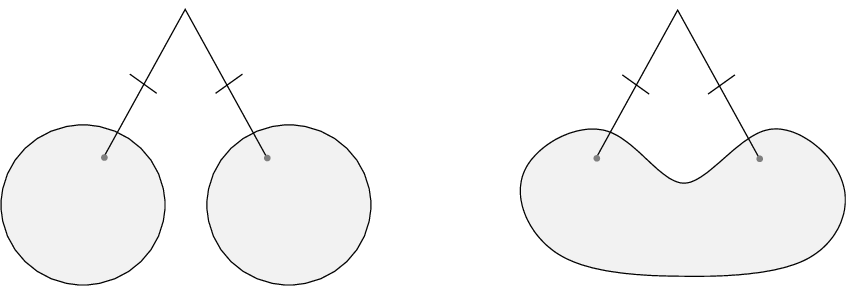}}
%%
%Next sentence care of Wolfram's Appendix
Because our two vertex operators in the bi-local term carry
momentum $k$ and $-k$ respectively, they occur precisely in the
same way as the propagator for the corresponding low-energy
field, and thus the effect of the deformation is to shift
the propagator:
\eqn\fullpropT{
G_2(k;T_c)\to
{i(1+F(k))\over k^2+i\epsilon}\biggl(e^{T_c(k^2+i\epsilon)}
- e^{T_0(k^2+i\epsilon)}\biggr).
}
In terms of the $\mu$ cutoff, the full
momentum-space propagator is parameterized as
\eqn\fullprop{
{i(1+F(k)) \over k^2-\mu^2+i\epsilon}.
}

In identifying our deformation with a shift in the propagator,
we have not implemented any extra subtraction prescription (such as
normal ordering) to remove divergences when the $V^{(k)}$
approach the $V^{(-k)}$.  As we will see in detail in \S3, this
divergence integrates to zero once we regulate
the theory and does not require any such
subtraction procedure.  (That is, in the field theoretic
organization of the string diagrams which
we are using \ZwiebachIE\PolchinskiRQ,
all such divergences arise in the propagator, which we have
regulated.)

$F(k)$ is constrained
as follows.
\item{1.} In order to preserve conformal invariance of the
worldsheet theory, we
demand that $F(k)$ vanish when $k$ is off-shell.
\item{2.} In order to cancel the divergences
coming from tadpoles, we need \eqn\condIF{ F(0)=-1+{\cal O}({1\over
T_c}), } and in order to precisely cancel the zero mode propagator,
we will require \eqn\Fnocorr{ F(0)=-1. } The latter condition
ensures that we remove the full zero mode propagator from the
tadpole contribution, rather than leaving behind a contribution
scaling like an extra massive tadpole as would occur if we kept a
nontrivial ${\cal O}({1\over T_c})$ contribution allowed by
\condIF.
%\item{3.} In order to preserve the cutting rules for unitarity,
%we expect that $F(k)$ must be real for real $k$
%and have no poles or other
%singularities in one of the half $k^0$ planes (say
%the lower half plane for definiteness). (As we will see, in the
%simplest version of $F(k)$ that we will study, unitarity
%is ensured more trivially though we will implement this condition
%nonetheless.)
%\item{4.} In addition, we would like the deformation
%\basicdefmu\ to
%contribute perturbatively to otherwise nondivergent
%diagrams, \ie intoduce no new divergences.
\item{3.}  We require $F(k)$ to be consistent with unitarity of
the resulting perturbative S matrix.  The simplest way to ensure
this, which we will employ here, is to choose an $F(k)$ such that
the deformation of the propagator does not contribute except in
precisely cancelling the tadpole contributions, leaving behind
tadpole-free diagrams which satisfy the cutting rules.

\noindent
One choice of $F$ we have found consistent with the criteria 1 --
3 is \eqn\simpleF{F(k)=\lim_{\eta\to 0} F_\eta(k)=\lim_{\eta\to
0}{\eta^2\over (k^0+|\vec k|-i\eta)(k^0- |\vec k|-i\eta)}. }

%(MAYBE WE WANT TO INCLUDE OTHER "SIMPLE" POSSIBILITIES FOR F
%SATISFYING 1-3 SUCH AS $-\eta^d\delta_\eta^{(d)}(k)$).

In a tadpole diagram, \basicdefmu\TNLST\ appears integrated with the
energy-momentum conserving delta function $\delta^d(k)$ for the
propagator in the tadpole part of the diagram. This picks out the
integrand evaluated at $k=0$, for which the factor \simpleF\
becomes ${\eta^2\over (-i\eta)^2}=- 1$. The entire propagator
strictly at $k=0$ is then (in the Schwinger parameterization)
\eqn\cancelledpropT{ G_2(k= 0)=\lim_{T_c\to\infty, T_0\to 0}
(1-1)(T_c-T_0) = 0, } or, in the massive QFT regularization
scheme, \eqn\cancelledprop{ (1-1){i\over -\mu^2+i\epsilon} = 0, }
as depicted in \deadants. Again, we refer to this mechanism for
decoupling the zero mode as decapitating the tadpole.

$F_\eta(k)$ in \simpleF\ can be written as
\eqn\expandF{
F_\eta(k)=\pi^2 (k^0+|\vec k|+i\eta)(k^0-|\vec
k|+i\eta)\delta_\eta(k^0+|\vec k|)\delta_\eta(k^0-|\vec k|)
}
where $\delta_\eta(x)={1 \over \pi}
{\eta\over{x^2+\eta^2}}$ is a regulated
Dirac delta distribution.
As such, when $F(k)$ is integrated against a smooth function,
% at $k\to 0$,
it vanishes.
%(actually it does not contribute unless
%integrated against a distribution as singular as $\delta^d(k)$).
As we have seen, when integrated
against $\delta^d(k)$ (which is of course not smooth at $k=0$)
it is $-1$, so that the deformation cancels the tadpole
divergence.  We will see that these properties
of $F(k)$ imply that its only contribution to physical
S-matrix elements (ones at generic external momenta or set up
as scattering amplitudes of smooth wavepackets) is precisely
its cancellation of the tadpole divergences.

Although we will work with the specific form \simpleF\ for
$F(k)$, any choice satisfying criteria 1--3 is suitable.
Any such $F$ which preserves Lorentz symmetry
will give an identical perturbative S matrix,
so any parameters involved in this choice are not physical,
at least perturbatively.

%In some computations, $i\eta$ plays a very similar role to the
%familiar $i\epsilon$ in Feynman diagrammatics. Namely, it is a way
%of formally encoding a pole prescription for the evaluation of
%contour integrals which arise in loop diagrams.

We will perform computations with the following order of limits:
we first send $\epsilon\to 0$ and $\eta\to 0$, then remove our IR
regulator by taking $T_c\to\infty$ (alternatively, $\mu\to 0$).
The $\epsilon\to 0$ and $\eta\to 0$ prescriptions are applied
integral by integral, diagram by diagram (\ie\ these limits are
taken before summing over infinite series of diagrams).
%Both
%$\epsilon$ and $\eta$ will play
%the role of picking out appropriate poles in loop integrals,
%and therefore are set to zero first, while $\mu$ or
%$T_c$ is an IR
%regulator needed to define the loop integrals for massless
%particles.
We refer to this regularization scheme as {\it the
padded room}.

This prescription involves two minor subtleties. Before taking
$\eta\to 0$, our deformation \basicdefmu\TNLST\ includes off-
shell (non-BRST invariant) vertex operators $V^{(\pm k)}$ with
$k^2\ne 0$.  Calculating the effects of our deformation
perturbatively, as we are doing, thus involves diagrams with
insertions of off-shell vertex operators.  In tadpole diagrams,
energy-momentum conservation projects the deformation onto $k=0$
so this issue does not arise.  In other diagrams, we need to
define our prescription and check that the non gauge-invariant
contributions vanish as $\eta\to 0$ (the limit we are taking in
which $F(k)$ has support only at $k=0$).  Our prescription for the
finite $\eta$ theory before taking the limit $\eta\to 0$ is to
work in a specific gauge (fixing the worldsheet metric up to
moduli to be integrated over) and calculate correlation functions
of the (on-shell and off-shell) vertex operators in  the
worldsheet CFT on this Riemann surface as in \PolchinskiJQ. We
will see in \S3  that the integration over $k$ in
\basicdefmu\TNLST\ involves $F(k)$ convolved with a smooth
integrand in the regulated theory, so that the
deformation makes a vanishing contribution as $\eta\to 0$. This
will depend simply on the local behavior of the $V^{(\pm k)}$ near
other vertex operators and degenerations of the surface.

In regulating the theory to produce a finite integral
over $k$ for general diagrams, note that a UV
regulator is also important in intermediate
steps of the calculation. For finite
$\eta$, the wedge propagator scales as $\eta^2/k^4$ for large $k$,
%%$${i(1+F_\eta)\over k^2-\mu^2+i\epsilon}\sim{ i\eta^2\over k^4}$$
(in the UV), which is not soft enough to prevent UV divergences in
the diagrams we are adding with wedge propagators in loops.  These
must be regulated. Once we regulate in the UV, all such diagrams
are proportional to (positive powers of) $\eta$, and these terms
all vanish diagram-by-diagram in the UV regulated theory once we
impose our limit ($\eta\to 0$).  In the Schwinger
parameterization, we can regularize in the UV with our parameter
$T_0$ in computations in which the loop integrals are defined by
Euclidean continuation.\foot{In rotating from Lorentzian to
Euclidean loop momentum integrals, an extra pole must be included
from \simpleF; however this pole does not contribute anything in
our regulated theory, as will become clear in \S3.} Alternatively
we can simply cut off the $k$ integrals at some scale $M_{UV}$. We
will see in \S3\ that all such loop contributions will vanish
regardless of the details of the choice of UV regulator.  (Note
that in the tadpole diagrams, the UV behavior is irrelevant since
the momentum $k$ is strictly zero.)

%fact completely UV finite and independent of the UV regulator. Due
%to the soft UV behavior of string loops (in which $T_0$ provides a
%physical UV regulator), this techincal subtlty deos not obtain in
%string theoretic computations\foot{Of course, since our
%deformation is not manifestly modular invariant, it is not clear
%that our deformation should appear with a finite $T_0$. However,
%trating $T_0$ as in the undeformed case is the most natural
%choice, and it works, so this will be our convention of choice.}.

This prescription \Fnocorr\cancelledpropT\cancelledprop\ for
cancelling divergences caused by
radiative tadpoles is reminiscent of the
prescription for renormalization of UV divergences
via counterterms in quantum
field theory.  Although our deformation has a large effect in
cancelling the divergences from tadpoles, it can be treated
perturbatively via (stringy) Feynman diagrams much like
counterterms in quantum field theory.  In both cases,
the (infinite) corrections appear in one to one correspondence
with divergences in the uncorrected theory, cancelling them
precisely.

%-----------------------[ Radiative Corrections ]--------------------%
\subsec{Radiative corrections: stability and moduli masses}

It is important to ask whether the specific form of $F(k)$
required by the criteria of the previous subsection is
preserved by loop corrections.  By construction it is immediate
that loop corrections to the ``head'' of the tadpole do not affect
the decapitation, which occurs at the level of the ``neck'' (\ie\
at the level of the propagator, regardless of
the form of the one-point amplitude to which it attaches).

In fact, loop corrections to the propagator itself also preserve
the cancellation of divergences. To see that this is the case,
take the 1PI self-energy, $\Sigma$, and use it to correct the
propagator including the modification \cancelledprop\ in  the
tree-level propagator.  One finds (for example in the field
theoretic regularization scheme)
\eqn\geometricseries{
G_{2,Ren}(k;\mu) = { 1+F \over k^2-\mu^2} \left(
1+\Sigma { 1+F \over k^2-\mu^2} +
\left( \Sigma {1+F \over k^2-\mu^2} \right )^2 + \cdots \right)
={1+F \over k^2-\mu^2 -(1+F)\Sigma} .
}
The fact that $1 + F(k)$ remains in the numerator
of the corrected propagator
clearly shows that the cancellation persists at zero
momentum and the renormalization of the propagator does
not change the fact that the tadpoles (now with renormalized
propagator for the neck) are decapitated.

Furthermore, this exhibits the following important physical
feature of our construction.  Nonzero modes in \geometricseries\
are not affected by $F(k)$, and are subject to generic mass
renormalizations included in the quantum self-energy $\Sigma$.
For models in which this renormalization produces positive mass
squared for all the scalars (\ie\ models in which the second
derivative of the effective potential is positive in all
directions about the starting value), the fluctuating modes of the
moduli are lifted!  One example of this is the $O(16)\times O(16)$
heterotic string, whose one-loop potential energy in Einstein
frame is proportional to $+e^{(5/2)\Phi}$.
Another example would be a pair of D-branes with
a repulsive force between them.

On the other hand, there are models in which some of the moduli
have negative mass squared at one loop, leading to tachyonic
instabilities for nonzero modes.  The resulting striped phases
may be interesting to study, but for now let us discard these
cases since these instabilities will drive us away from the
simplest case of Poincar\'e invariant flat space.  Examples of
this latter class of 1-loop tachyonic backgrounds include
Scherk-Schwarz compactifications and D-brane--anti-D-brane
systems.

In this analysis it is important to follow the padded room
regularization prescription specifying that the limit $\eta\to 0$
be taken diagram by diagram. In particular, for finite $\eta$, the
right hand side of \geometricseries\ has poles in the complex $k$
plane corresponding to solutions of the linearized field equations
with exponential growth along the spacetime coordinates
$x^\mu$.\foot{We thank the authors of \IRgroup\ and N. Kaloper
and E. Martinec
for emphasizing this issue to us.} As we take $\eta\to 0$, these
solutions revert to oscillating solutions; summing the resulting
diagrams then gives the finite result above.
If instead we were to sum these diagrams before taking $\eta\to 0$,
thereby studying the RHS of \geometricseries\ first at finite
$\eta$, we would expect divergences arising from these exponentially
growing solutions (similar to divergences caused by tachyons in
loop diagrams).
% which would be ill-behaved when we finally take
%$\eta\to 0$.
Importantly, this order of operations is explicitly
disallowed in our regularization prescription; the limit $\eta\to
0$ is part of the definition of each diagram
%, specifying a pole prescription,
and must be taken before doing the sum in
\geometricseries.
%Thus no divergences or ambiguities arise in the
%limit $i\eta\to 0$.
% THE FOLLOWING IS ONLY TRUE FOR OUR SPECIAL FORM OF F(k),
% WHILE THE ABOVE IS GENERAL, HENCE I COMMENTED IT OUT....
In fact, as we will see in \S3, diagram by diagram our deformation
does not contribute in loop propagators; $F(k)$ integrated against
the rest of the amplitude vanishes unambiguously, diagram by
diagram.

A related issue is the question of whether nonperturbatively
the decapitated theory has other background solutions, different
from flat space, with consistent (in particular, unitary) physics.
(For example, in the presence of our deformation, could one still
start with a solution in scalar field theory with the scalar field
rolling down the potential hill and expand around this solution to
produce a consistent theory?)
If there exist other solutions which are in fact connected
physically to our flat space solution, it
would be interesting to study nonperturbative dynamics that
may select which background will arise naturally when this framework
is considered in a cosmological context.
This very interesting
question we leave for future work.

%-----------------------[ Decapitatating Gravity ]--------------------%
\subsec{Decapitating the graviton tadpole}

We so far formulated our deformation for massless scalar fields.
The tadpole generated for the (trace of the) graviton is the
cosmological constant and is of particular interest.\foot{We could
restrict our attention to scalars by considering tadpoles for
the scalars
arising in the open string sector on D-branes with broken
supersymmetry  (see the next subsection).}

Since the graviton tadpole (cosmological constant) is one of
the main motivations for pursuing this direction, we wish to
generalize our prescription to a modification of the graviton
propagator which cancels its zero mode.
In particular, for the procedure under discussion to
be useful in a simple closed string example (like the $O(16)
\times O(16)$ heterotic string)
we need to decapitate the graviton also so as to avoid
generating large curvature.

It may also be interesting in
some circumstances to decapitate the scalars but shift the
gravity background in the standard way to obtain dS or AdS
space.
%To this end, it is perhaps worth emphasizing that the decapitation
%of tadpole-driven divergences does not preclude additional shifts
%of the scalar VEVs, and, in particular, of the graviton trace.
That said, we content ourselves in the following to the most
simple case of asymptotically flat space, leaving generalizations
to future work.

In expanding about flat space, Lorentz invariance implies that
the only tadpole contribution from the gravitational sector
comes from the trace of the graviton.  The trace can be
gauged away for nonzero momentum, but at zero momentum the
gauge transformation required to do so would not vanish at
infinity.  The worldsheet manifestation of this is the presence
of an extra BRST-invariant vertex operator
at zero momentum transforming as a spacetime scalar,
which we will denote by $V_{tr G}^{(k=0)}$.
(This mode is degenerate with but independent from the
zero-momentum mode of the dilaton.)  Defining
\eqn\gravop{
V^{(k)}_{tr G}\equiv : V_{tr G}^{(k=0)} e^{ikX}:
}
we add to the worldsheet action
\eqn\defgrav{
\delta S_{ws}^{G}=\int {d^dk\over (2\pi)^d}
{F(k)\over k^2+ i\epsilon}\biggl(e^{T_c(k^2+i\epsilon)}-
e^{T_0(k^2+i\epsilon)}\biggr)\int V^{(k)}_{tr G}
\int V^{(-k)}_{tr G}
}
As in the case of the scalar fields, before taking $\eta\to 0$
this involves off-shell vertex operators $ V^{(\pm k)}_{tr G} $ with
$k\ne 0$ included in \defgrav.  Again, we can compute in a
fixed gauge and show that these contributions vanish when
$\eta\to 0$.

%The generalization is as follows:  look at the usual
%momentum space graviton propagor in superstring theory (in an
%appropriate gauge),
%call it $G^{\mu \nu \rho \sigma} (k)$.  Then the analogue of
%\basicdef\ is
%\eqn\basicdefgrav{\delta S_{ws} =
%\int dk \int d\sigma_1 \int d\sigma_2 F(k)
%G^{\mu \nu \rho \sigma} (k)
%V_{\mu \nu}(k, \sigma_1) V_{\rho \sigma}(-k, \sigma_2)
%}
%where $V_{\mu \nu}$ is the graviton vertex operator.
%This is just like in the scalar case; here $G^{\mu \nu
%\rho\sigma}(k)$
%plays the role of the usual $1 \over k^2 + i \epsilon$ scalar
%propagator.
%According to Nemanja, in de Donder gauge, this
%$G^{\mu\nu\rho\sigma}$
%is of the form
%\eqn\Gform{G^{\mu\nu\rho\sigma} (k) = {\CP^{\mu \nu \rho
%\sigma} \over k^2 + i \epsilon}.}
%where the tensor $\CP$ is independent of the momentum.

%As in the case of the scalar field, we will regulate this in
%the IR by cutting off the modular parameter $T$ in the string
%propagator at some large value $T_c$ to be taken to infinity
%at the end of the computations.  This means again that we
%replace the ${1\over k^2+i\epsilon}$ in \defgrav\ by ${1\over
%k^2+i\epsilon}
%(e^{T_0 (k^2+i\epsilon)} -e^{T_c (k^2+i\epsilon)})$ to obtain
%the analogue of \TNLST.

As in the scalar case, this suffices to cancel all tadpole
divergences at any loop order. (Note that in contrast to the
scalar case, the self-energy of the graviton of course does not
include a mass by gauge invariance.)  Since we only modified
the zero mode of the graviton, we do not expect problems with
gauge (diffeomorphism) invariance to be introduced by our
prescription; gauge transformations which die at infinity
cannot act on the strict zero mode of the graviton.  Acting
only on the zero mode also ensures that the graviton
responds to ordinary local sources of stress-energy in the
usual way, as we will exhibit for the S-matrix in \S3.

%-----------------------[ Open String Examples ]--------------------%
\subsec{Open string examples}

It is worth emphasizing that we may consider tadpoles for
scalars independently
of gravitons by considering a non-supersymmetric combination of
D-branes in a supersymmetric bulk theory.  In such a situation,
any closed string tadpoles can be absorbed in radial variation of
the fields (if the D-branes are at sufficiently high codimension).
In order to produce an S- matrix with positive mass squared for
the nonzero modes of the scalars, we can for example choose a pair
of branes which repel each other at long distance.
% (such as theD0-D6 system).
(Note that we may not choose an attractive
potential $V( r)\sim -{1\over r^n}$ such as arises in a simple
D-brane-anti-D-brane system since $V^{\prime\prime}( r)<0$ in that
case; we can instead choose a repulsive potential $V( r)\sim
+{1\over r^n}$ which has $V^{\prime\prime}( r)>0$.)  In such a
system we may decapitate the tadpoles for scalars on one or both
of the branes (shifting the nondecapitated fields to the
appropriate time-dependent solutions describing motion of the
corresponding brane).

%-----------------------[ BRST Analysis]--------------------%
\subsec{BRST analysis}

In \PolchinskiJQ\LaXK, the loop corrected equations of
motion for massless fields were derived by requiring
that BRST trivial modes decouple from string S-matrix elements.
%in the presence of tadpoles.
One considers a diagram
\ifig\ants{Before decapitation, the tadpole spoils the decoupling
of BRST trivial modes.  Decapitation adds a diagram precisely
cancelling {\bf only the divergent part of} this anomaly: see text.}
{\epsfxsize3.0in\epsfbox{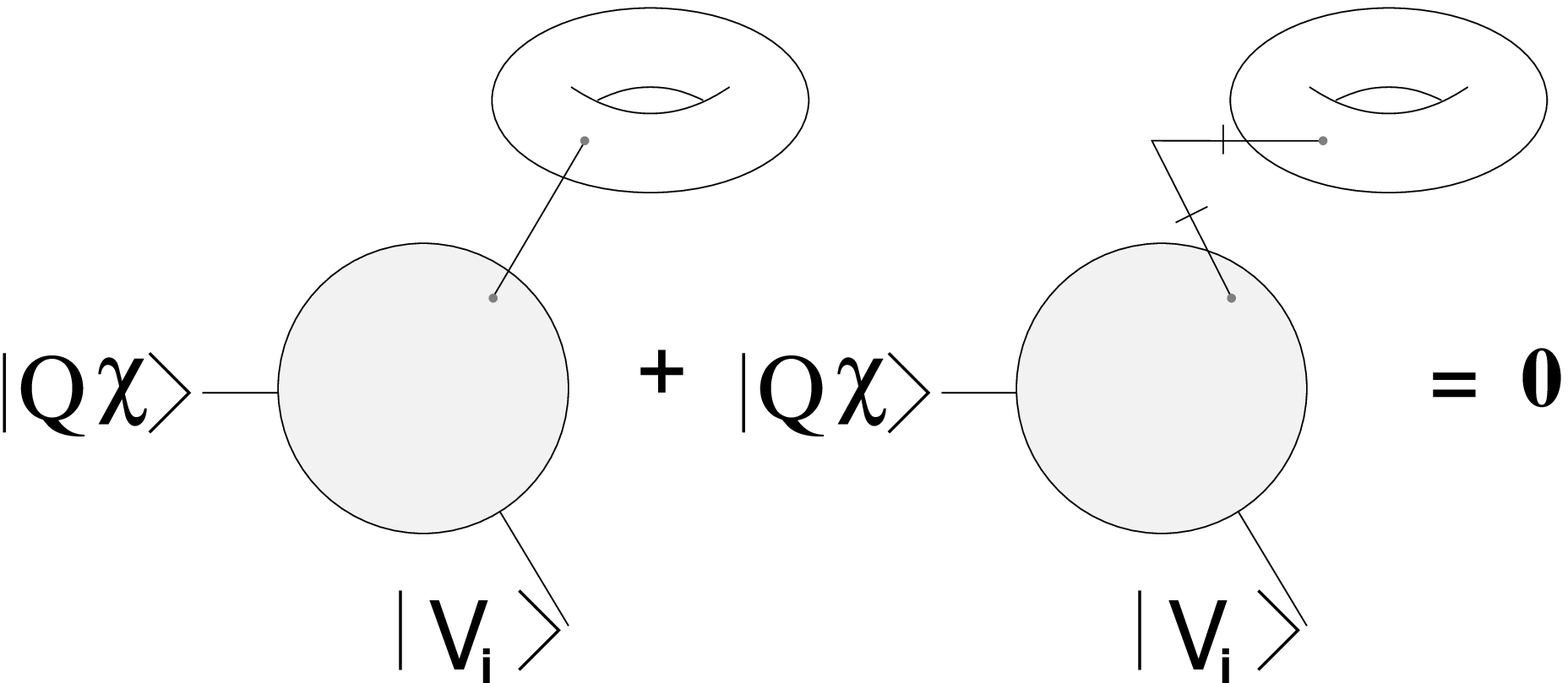}}
\noindent with one BRST trivial vertex operator
$Q_B\chi\equiv \oint_z j_B \chi(z,\bar z)$ and
any number of physical vertex operators $V_i$.
One can deform the contour of integration
away from $\chi(z,\bar z)$ so that
the BRST operator $Q_B$ acts on the other insertions
in the diagram.  $Q_B$ kills the remaining (physical)
vertex operators.
{\it On a degenerating tadpole neck, it contributes
a finite anomalous piece which our deformation as it stands
does not cancel.  A modification aimed at correcting this
is in progress.}

%-----------------------[ Spacetime Description ]--------------------%
\subsec{Effective field theory description}

A useful heuristic way to describe our prescription is to
consider the momentum-space effective action for a scalar field
$\phi$ whose tadpole we are decapitating.  The presence
of the discontinuous object $F(k)=\lim_{\eta\to 0}F_\eta (k)$
complicates the analysis of the field theory (the limit
$\eta\to 0$ being taken diagram by diagram in the
S matrix as we explained in \S2.1\S2.2).  We will
ignore all such subtleties in this subsection with
the aim of gaining some further intuition for
the physics of the deformation.  Taking into account
the modification we have made to the propagator, this
effective action is
\eqn\scalact{
\int d^dk \biggl[\phi(k) \bigl({{k^2-\mu^2}\over 1+F(k)}-
\lambda_2\bigr)
\phi(-k)\biggr]-\lambda_1\phi(0)-\int d^dk \int d^dk^\prime
\lambda_3\phi(k)\phi(k^\prime)\phi(-k-k^\prime)-\dots
}

%Ignoring the cubic and higher terms in the effective potential
%(i.e. assuming $\lambda_n\sim 0, n>2$),
This leads to the equation of motion
\eqn\eqphi{
\phi(-k)=\lambda_1\delta^d(k){1\over{{{k^2-\mu^2}\over 1+F(k)}}
-\lambda_2}}
plus subleading terms involving the higher ($\lambda_{n>2}$)
terms in the effective potential.
Because of the $F(0)=-1$ contribution, the right hand side here
is of the form $f(y)\delta(y)$ with $f(0)=0$, so this vanishes.
That is, $\phi(0)$ is not forced to shift by the tadpole once
we include our modification of the kinetic term (corresponding
to our original modification of the propagator).

This description involving a nonlocally modified action
may be useful but we will mostly stick to the S-matrix
formalism (natural in perturbative string theory) we have been
developing.

%-----------------------[ In Contrast ]--------------------%
\subsec{In contrast}

Before returning to the S-matrix description,
it is worth noting at this point that
our prescription is different from two somewhat similar
manipulations that might be confused with it.

%-----------------------[ Boundary Conditions ]--------------------%
\subsubsec{Removing the zero mode by boundary conditions}

First, in field theory one might consider removing the zero mode
of a massless field by putting the system in a box with
appropriate boundary conditions.  For example, consider a scalar
field with a tadpole (say a linear potential) in a box.
Imposing Dirichlet boundary conditions removes the
zero mode.
However, since this
does not change the basic equation of motion,
half of the remaining modes still respond to the
linear term in the potential.  Adiabatically decompactifying
the box therefore leads to an unstable theory.

Decapitation works not by selecting particular
solutions of the original equation of motion, but by changing
the equations of motion.
In our case \scalact, there is no linear term for nonzero modes,
and hence no instability left in the system once we remove the
zero mode by our decapitation prescription.  Also, our
analysis of decapitation involves a regulation prescription
compatible with an S matrix description, whereas introducing
a box as an IR regulator would not have this feature.

%-----------------------[ Stringy IR Mods ]--------------------%
\subsubsec{String IR modifications}

As discussed in \AharonyPA, the bilocal deformation
$\delta S\sim \int V\int V$ can be obtained by deforming the
action locally by
\eqn\deflam{
\delta S=\int d^2z \lambda V
}
and integrating over $\lambda$ with a Gaussian weight.

Recently a modification of string theory has been proposed in
\FriedanAA\ which involves considering fluctuating couplings
$\lambda$ on the worldsheet.  In our case \deflam, $\lambda(k)$ is
a constant on the worldsheet, whereas in \FriedanAA, $\lambda=\lambda(z,\bar z)$
is a fast varying function of the worldsheet coordinates,
and in particular explicitly does not include a
worldsheet zero mode.

%-----------------------[ UNITARITY ]--------------------%
\newsec{Effects of deformation on general diagrams and
unitarity}

We have so far established that our modification removes the
tadpole divergences associated with massless fields.  We must
now address the question of what other effects the modification
has, and in particular determine whether the S-matrix resulting
from our deformation is unitary.

Because of the simplicity of the $F(k)$ we chose for our
deformation, we will see in fact that it does not contribute to
physical S-matrix amplitudes beyond its cancellation of tadpole
divergences, and that unitarity is therefore satisfied.

In particular, as we have seen, $F(k)$ vanishes when integrated
against any smooth function (its nonvanishing
contribution cancelling the tadpole arises from its
integration against a delta function $\delta^d(k)$).  The
question is then whether the $k$-dependence of the integrand in
amplitudes obtained by bringing down powers of \TNLST\ and
\defgrav\ is sufficiently smooth,
modulo (non-smooth) $\delta^d(k)$ factors
coming from tadpole contributions.  (Note that we are working
with a UV cutoff which ensures no divergence from the UV
end of the $k$ integration.)
Generic diagrams involving smooth
wavepackets integrated over external momenta as well as
ordinary loop momentum integrals indeed turn out to have this
property in the padded room, \ie\ in our regularization
prescription.

Thus we are interested in the $k$-dependence of amplitudes with
insertions of $V^{(\pm k)}$,  near potential singularities in
the integrand.
The $V^{(\pm k)}$ can be slightly off shell before we take the
limit $\eta\to 0$, and we define their amplitudes by working in
a gauge-fixed worldsheet path integral.  The possible
singularities in the integrand arise as the $V^{(\pm k)}$
approach other vertex operators $V^{(p_i)}$ or degenerating
internal lines carrying momenum $p_i$.  In both cases, the
behavior is determined locally on the Riemann surface and has
the structure $\int {{d^2z}\over{|z|^{2+2k\cdot p_i}}}\sim
{1\over (k+p_i)^2-m_i^2+i\epsilon}$.  As in the UV,
these potential divergences
are cut off in the IR by our regularization prescription.

%-----------------------[ Non 1PI Contributions]--------------------%
\subsec{Non-1PI contributions}

Let us consider first diagrams for which cutting
an $F(k)$ contribution to the propagator (which
we will refer to as a ``wedge propagator''
contribution) breaks the diagram in two.  For
this non-1PI propagator there are two cases.  One
is what we have already accounted for: the wedge
propagator attaches to the head of a tadpole
(with no incoming momentum); in this case the
wedge contribution cancels the divergence from
the tadpole (in fact the whole massless
propagator contribution) by construction.  The
second case is that the the wedge propagator in
question connects to a subdiagram with incoming
momenta $q_{i}$, so that generically there is
nonzero momentum
$k\equiv \sum_{i=1}^n q_i$ flowing through the
wedge propagator.

At generic incoming momentum, since $k\equiv \sum_{i=1}^n q_i \ne
0$, $F(k)$ does not contribute (since $F(k\ne 0)=0$).  Similarly,
if we consider a smooth wavepacket in the incoming momenta $q_i$,
the relevant part of the amplitude is \eqn\treewp{ \int \prod_i
d^dq_i f(q_i) F(\sum_i q_i) {i\over (\sum_i q_i)^2+i\epsilon}
\biggl(e^{[(\sum_i q_i)^2+i\epsilon]T_0} - e^{[(\sum_i
q_i)^2+i\epsilon]T_c}\biggr) } We can change basis in the $q_i$ to
obtain an integral over $\sum_i q_i$ (the argument of $F$ in this
amplitude); it is then clear that the integrand is sufficiently
smooth at
$\sum_i q_i=0$ and because of the convolution with $F$ this
amplitude vanishes.\foot{In fact for normalizable wavepackets
$f(q_i)$ is not only smooth at $\sum_i q_i=0$  but vanishes
there.}

%-----------------------[ Forces 'tween Branes]--------------------%
\subsubsec{Forces between D-branes}

One type of one-particle reducible diagram of particular interest is
that describing the force between D-branes, so
let us study this explicitly.  Here we have at
leading order
\ifig\ants{Diagrams contributing to the force between parallel
branes.}
{\epsfxsize3.0in\epsfbox{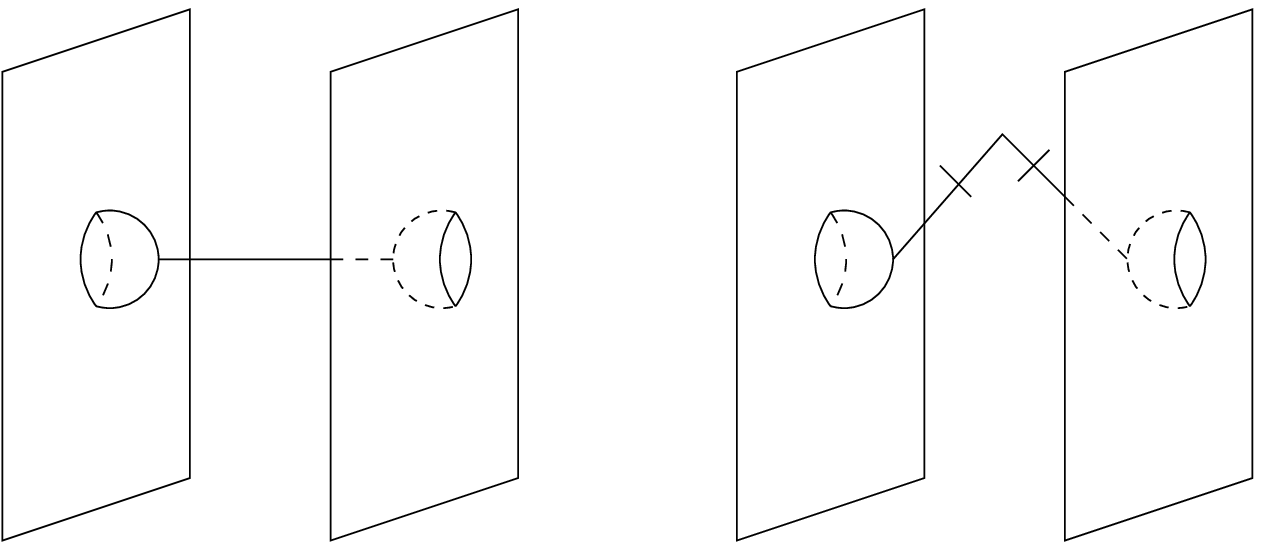}}
The correction term is proportional to
\eqn\Dcorr{
\int d^n \vec k_{\perp} {{F(\vec k_\perp)}\over
\vec k_\perp^2 +i\epsilon}\biggl(e^{-T_c(\vec
k_\perp^2 +i\epsilon)}-
e^{-T_0(\vec k_\perp^2 +i\epsilon)}\biggr)
}
where $\vec k_\perp$ denotes the momenta in the
$n$ transverse directions to the D-brane.  This
contribution vanishes, as can be seen by plugging
in the above expression for $F$ in terms of delta
functions \expandF.
So as expected from the general arguments above,
we see explicitly here that the force between
gravitational sources such as D-branes is not
changed by our decapitation of the tadpoles of
the closed strings exchanged.

%-----------------------[ 1PI Contributinos ]--------------------%
\subsec{1PI contributions}

Consider a general contribution involving wedges
which carry loop momentum (\ie\ a diagram which
is 1PI
with respect to cutting at least some of the
wedges).  We would like to know
if this diagram is nonzero (and if it is nonzero,
we
would like to know if it
preserves unitarity of the S-matrix).

Let us focus on one wedge at a time, with
momentum $k$.
If the Riemann surfaces are smooth and the vertex
operators
are separated from each other and from the
$V^{(k)}$'s, then
the integrand will be nonsingular.  The potential
divergences as $k$ varies
come from the degenerations of the Riemann
surface approaching the $V^{(\pm k)}$'s and/or
the approach of vertex operators to each other.
These
can always be viewed as IR divergences or poles
in the
S-matrix.
So we can focus on the region of the moduli space
of the Riemann surface near IR limits and poles.  (Again,
note that any UV divergences are cut off.)

Using this, the structure of the potentially singular part of the
$k$-dependent integrand in the amplitude is
\eqn\kpart{\eqalign{\int d^dk & F(k){i\over k^2+i\epsilon}
(e^{T_c( k^2+i\epsilon)}-e^{T_0( k^2+i\epsilon)})\cr & \prod_i\int
d^dp_i f(p_i){1\over (k+p_i)^2-m_i^2+i\epsilon} (e^{T_c(
(k+p_i)^2-m_i^2+i\epsilon)}-e^{T_0(
(k+p_i)^2-m_i^2+i\epsilon)}),\cr } }
%\eqn\genstruc{
%{\cal K}\sim \prod_{i=1}^n{i\over {(k+p_i)^2-
%m_i^2-\mu^2+i\epsilon}}
%}
times a factor of $T_c$ if the $V^{(\pm k)}$ approach each other
(\cf\ \Gcutoff).
Here the $p_i$ are linear combinations of some subset of the
momenta (including in general both loop and external momenta).
That is, the propagators in \kpart\ come from pieces of the
diagram in which a $V^{(\pm k)}$ line hits a line carrying
momentum $p_i$. In the case that $p_i$ is a linear combination of
external momenta, then we take the function $f(p_i)$ to be a
nontrivial smooth wavepacket.\foot{This wavepacket should die
off fast enough for large momentum so as not to introduce
new UV divergences; we may in any case include a UV cutoff
$M_{UV}$ on the external momentum integrals as well as
on the internal ones.}  When $p_i$ involves a loop
momentum, then $f(p_i)$ encodes any further momentum dependence in
the amplitude beyond the pole contribution, and again is a smooth
function.

As before, whether this
contribution survives is determined by whether
the integrand as a function of $k$ can become
singular
as $k$ varies.
This is clearly averted here since the
only singularities of the integrand are
the poles from the propagators,
and for finite
$T_c$, the
expansion of the exponentials for small
$(k+p_i)^2-m_i^2$
kills the factor of $(k+p_i)^2-m_i^2$ in the
denominator.
So we see that the $F$ terms do not contribute in
loop (1PI) propagators, just as we found for
non-1PI propagators in physical S-matrix amplitudes.

%-----------------------[ New Tadpoles (sic) ]--------------------%
\subsec{New tadpole diagrams which vanish}

It is worth mentioning that the tadpole
contributions formally include the following
diagrams introduced by our modification:

\ifig\extra{
0+(-0)=0.
}{\epsfxsize1.5in\epsfbox{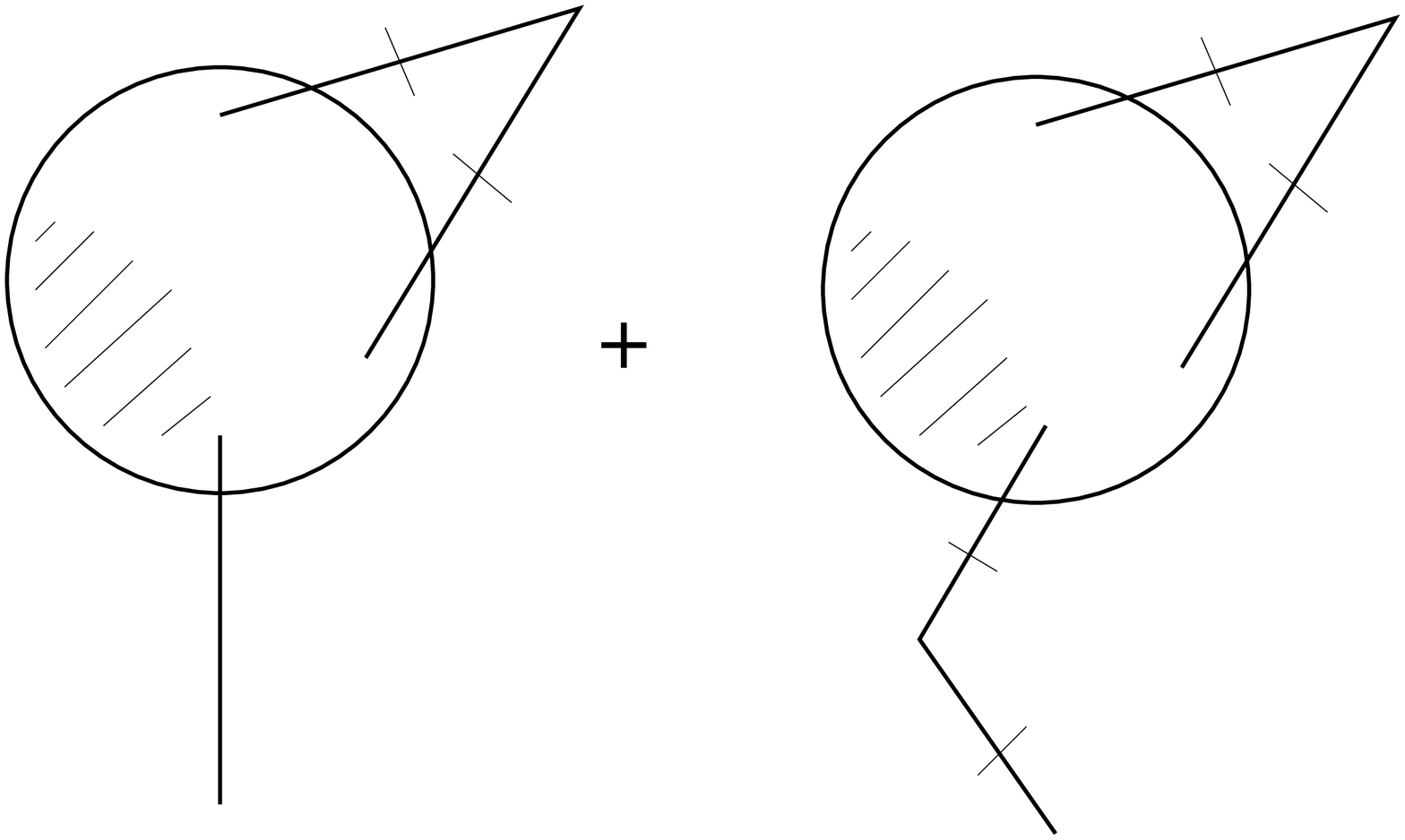}}

However, these diagrams cancel.
Not only do they cancel each other exactly via decapitation,
but they are
separately zero because as we have just derived, the $F$'s do
not contribute
in loops.  This is related to the comment
in \S2\ about the absence of a need for a
normal-ordering prescription for the product $V^{(k)}
V^{(-k)}$.

%-----------------------[ Explicit 1-Loop ]--------------------%
\subsec{Explicit evaluation at one loop}

The above general arguments suffice to establish that our
deformation proportional to $F$ does not contribute except in
decapitating the tadpoles.  It is nonetheless instructive to work
out explicitly a simple 1-loop example in quantum field theory to
illustrate the effect.

Let us consider a one-loop graph involving
two virtual massless scalar particles (whose tadpoles
we are decapitating) with total momentum $p$
running through it and loop momentum $k$.  This is given
by (up to an overall real constant)

\eqn\Iloop{
\lim_{\mu\to 0}\lim_{\eta\to 0}\int {d^dk\over (2\pi)^d}
{1+F_\eta(k)\over k^2-\mu^2+i\epsilon}
{1+F_\eta(k-p)\over (k-p)^2-\mu^2+i\epsilon}
}

Let us perform the $k^0$ integral by treating it as a contour
integral, closing the contour at infinity in the lower half
plane.
This is possible because the integrand falls off for large
$|k^0|$.
This picks up the residues of poles at
$k^0=\sqrt{\vec k^2+\mu^2-i\epsilon}$ and
$k^0=p^0+\sqrt{(\vec k-\vec p)^2+\mu^2-i\epsilon}$.
(Note that this follows even in the presence of the $F$ terms
because we constructed $F_\eta(k)$ to have no poles in the
lower half $k_0$ plane.)
Letting $E_k \equiv \sqrt{ |\vec k|^2 + \mu^2} $, this gives
\eqn\IloopII{\eqalign{
i\int {d^{d-1}\vec k\over (2\pi)^{d-1}} \biggl(
& {1+F_\eta(E_k-i\epsilon, \vec k)\over
2\sqrt{\vec k^2+\mu^2-i\epsilon}}
{1+F_\eta(E_k-i\epsilon - p^0, \vec k-\vec p)
\over (E_k-i\epsilon-p^0)^2-(\vec k-\vec p)^2-\mu^2
+i\epsilon}+\cr
& +{1+F_\eta(E_{k-p}-i\epsilon,\vec k-\vec p)
\over 2E_{k-p}-i\epsilon}
{1+F(p^0+E_{k-p}- i\epsilon,\vec k)
\over (p^0+E_{k-p}-i\epsilon)^2-\vec k^2-\mu^2
+i\epsilon}\biggr)\cr
}}
%This is
%\eqn\IloopIII{\eqalign{
%&i {\omega_{d-2}\over 2 (2\pi)^{d-1} }
%\int_{-1}^1 d\cos \theta \int_\mu^\infty dE_k \left( E_k^2 -
%\mu^2 \right)^{(d-3)/2}\cr
%&\left(
%(1 + F_\eta ( E_k, \vec k))  {1 + F_\eta( E_k - p^0, \vec k -
%\vec p) \over
%\mu^2 - 2 ( E_k E_p - |\vec p|| \vec k| \cos \theta ) }  +
%\right. \cr
%&\left. (1 + F_\eta ( p^0 + E_k, \vec k - \vec p ))  {1 +
%F_\eta( p^0 + E_k, \vec k ) \over
%\mu^2 +  2 ( E_p E_{p-k} + \vec p \cdot ( \vec p - \vec k) )}
%\mu^2 +  2 ( E_p E_{p-k} + |\vec p|^2 - |\vec p| |\vec k| \cos
%\theta )}
%\right)
%}}
%where $\omega_{d-2}$ is the volume of the unit $d-2$-sphere.
For generic external momentum $p$, the denominators in this
expression never vanish (for finite $\mu$, which is taken to
zero at the very end of the computation) where either of the
$F$ factors have support. Hence as argued for general diagrams
in the above subsections, here we see explicitly that the
deformation does not contribute in loop propagators.

%-----------------------[ Unitarity Revisited ]--------------------%
\subsec{Unitarity}

Because the $F$ terms do not contribute to amplitudes except in
cancelling massless tadpole contributions, we expect that
perturbative unitarity is satisfied.  This is manifest in simple
quantum field theories such as $\phi^3$ theory expanded
about $\phi_0=0$:  once the
diagrams including tadpoles are removed the remaining diagrams
satisfy the cutting rules for perturbative unitarity (see
figure 9).  This result
is clear also from the equivalence of the S matrix
resulting from decapitation and that obtained by
simply fine tuning away
the tadpole contribution order by order; the latter also removes the
tadpole diagrams leaving behind finite ones satisfying the cutting
rules.
\ifig\figUnitarity{
The one loop two-point function in decapitated $\phi^3$ theory.
(a) All diagrams with wedges in loops vanish identically.
(b) Decapitation ensures that all tadpole diagrams cancel.
(c) The remaining diagram respects the cutting rules by
construction.
}
{\epsfxsize3.0in\epsfbox{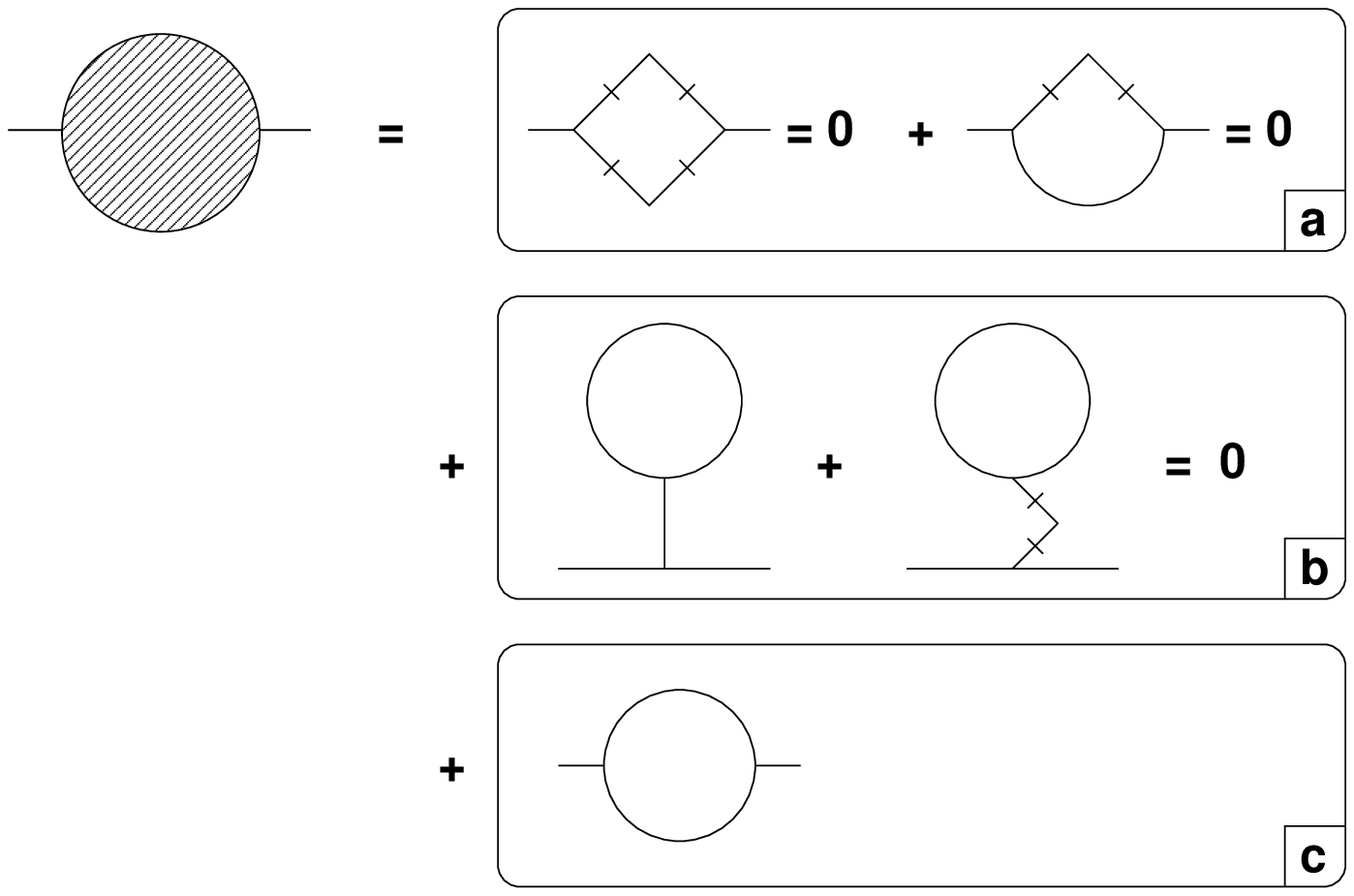}}

In ordinary bosonic string theory, one can formally argue for perturbative
unitarity by decomposing string diagrams into quantum field theory
diagrams made from propagators and hermitian vertices (the latter
containing
no boundaries of moduli space and therefore no poles) (see \eg\
\ZwiebachIE\ and the discussion in \PolchinskiRQ, chapter 9), and
then appealing to the field theory argument based on Cutkosky
rules. As we have shown in this section, the effect of our
deformation is precisely to cancel massless tadpole contributions
in this field theory language; the remaining diagrams satisfy the
cutting rules as usual if they do in the undeformed theory.  So if
the superstring perturbation theory works similarly to the bosonic
case in this regard, \ie\ if it is decomposable into diagrams,
formed from propagators and vertices, which satisfy the cutting
rules for unitarity (which seems plausible though it has not been
proved), then we can conclude that our deformation produces a
unitary theory.

Since the remaining diagrams describe forces that fall off
with distance, we expect cluster decomposition
to hold in our theories.  (This is again manifest in the perturbative
quantum field theory examples where the result is equivalent
to that one would obtain tuning away the tadpoles.)

Note that since we have shown that
tadpole-free diagrams are unaffected by our modification, the
analogous modifications of perturbative supersymmetric theories
would have no effect on the physical S matrix.
(An interesting future direction is to
apply our construction, perhaps field theoretically, to models
of low-energy supersymmetry with dynamical (nonperturbative)
supersymmetry breaking.)

%noncontribution for generic momenta (1-loop diagram
%explicitly)

%resulting unitarity modulo nonwavepacket, wrong order of
%limits nongeneric momenta

%-----------------------[ DISCUSSINO ]--------------------%
% its bosonic partner has been lifted, i suppose,
% this being a paper about nonsupersymmetric strings.

\newsec{Discussion}

Having argued for the unitarity of our S-matrix, let us now recap
and assemble the salient physical features of our system. Our
prescription leads to a class of unitary non-supersymmetric
perturbative S-matrices in flat space, parameterized by the VEVs
of the classical moduli, whose fluctuating modes are
generically lifted. We accomplished this by rendering non-dynamical
the zero modes of fields (moduli and the graviton) which would
otherwise be destabilized by tadpoles, via a modification of the
propagator for these fields in the deep infrared.  On the
worldsheet this modification arises as a perturbative NLST
deformation.  The tree-level S-matrix is the same as in the
unmodified theory; in particular the response of gravity to
localized sources of stress-energy is as in ordinary general
relativity and has not been removed by our mechanism.

The tadpoles in our examples are uniform over spacetime, and have
been effectively removed.  It is worth emphasizing that this is
not true of the cosmological term in the real world, which is
subject to phase transitions (variation in time) as well as
possible variation among different spatial domains.
%Even if this were not an issue,
Further, we have not so far identified a dynamical
mechanism for selecting our theory. In this regard, it will be
very interesting to study more systematically the space of
consistent IR deformations along these lines.

One result of our analysis which is in some sense disappointing is
the presence of parameters descending from the VEVs of the moduli
fields. Again, these arise because we can implement our
decapitation construction expanding about any point in the
classical moduli space having positive 1-loop quadratic terms in
the potential for all the moduli. The point in the moduli space
from which we start controls the couplings in the S matrix, while
the decapitation construction removes the tadpoles which would
otherwise generically drive the moduli away from the starting
point.  Our construction (for any choice of
$F(k)$ satisfying our criteria in \S2) does not entail any
parameters coming from $F(k)$, though it is possible that
more general choices of $F(k)$ that do affect non-tadpole
diagrams could also lead to consistent perturbative S matrices
in flat space or otherwise.

Continuous parameters are of course also seen in flat space SUSY
models with moduli spaces and in SUSY and non-SUSY versions of
(deformations of) the AdS/CFT correspondence (where the values of
the field theory couplings in the UV form a continuum of
parameters).  The novelty here is the persistence of
such a continuum
after supersymmetry breaking, in a background preserving maximal
(Poincar\'e) symmetry. (This also has something of an analogue in
known backgrounds--in flux compactifications even after
supersymmetry breaking, one has a finely spaced set of discrete
parameters which can allow one to effectively tune contributions
to the low-energy effective action, including the cosmological term
\BoussoXA\GiddingsYU\MaloneyRR\KachruNS\FreyHF.)

This work leaves open the possibility
that our perturbative string theories may not complete
to nonperturbatively consistent theories.  It was only
relatively recently that ordinary perturbative string
theories have been (in many cases) understood to
fit into a nonperturbative framework via string/M theory duality,
matrix theory, and AdS/CFT.  We do not have any concrete
results on this question;  perhaps something could
be learned by considering nonperturbative
features of decapitation in spontaneously broken
gauge theories.\foot{Work on a related question of
whether or not similar
modifications might be consistent
in the Higgs sector of the Standard Model is in
progress \higgsquestion.}   Also, it is possible
that the assumption we make about the undeformed superstring
diagrams satisfying perturbative unitarity relations
as in quantum field theory along the lines of the bosonic
case \ZwiebachIE\PolchinskiRQ\ is wrong because of
subtleties associated with superstring perturbation theory.
This loophole we find less plausible but in the absence
of a proof it certainly remains a possibility.

Although (as in the previously known cases listed above)
the parameters add to the lack of predictivity in
perturbative string theory,
there is a very appealing robust prediction in this class of
models.  Namely, our construction provides a mechanism
for solving the moduli problem, in that
for generic values of the parameters in our S-matrix, the
fluctuating modes of the moduli are lifted.

While in this paper we considered perturbative diagrams producing
tadpoles, our construction may also apply to situations in which
SUSY is broken dynamically at low energies. As an IR effect, we
can describe our modification in field theory terms, and
low-energy field theoretic SUSY breaking models may be amenable
also to such a deformation.  (Also, in some circumstances
classical SUSY breaking superpotentials may be dual to dynamical
ones.)

Similarly we may ask about non-flat backgrounds.
It will be interesting to consider whether we can decapitate
scalar tadpoles but not the graviton tadpole,
leading to a de~Sitter or anti
de Sitter solution.  It is also important to understand much
better the space of consistent string backgrounds,
in particular to understand how much fine tuning of
initial conditions is
required to land on the flat space backgrounds we have
exhibited in this paper.

Along similar lines, one may consider IR deformations of this sort
which involve different forms for $F(k)$.  In particular one can
imagine introducing a length scale $L$ above which the
decapitation acts nontrivially, rather than simply acting at zero
momentum.  As in \IRgroup, this may bring the approach closer to
applying to the real world cosmological term.

An important theme of this subject is the application of
renormalization ideas to infrared divergences.  Our prescription
here is analogous to renormalization via counterterms in that the
finite result we obtain arises from cancellation of quantities
that diverge as the cutoff is removed. It would be very
interesting to pursue the possibility of IR renormalization using
instead an analogue of Wilsonian renormalization involving
coarse-graining in momentum space.

\noindent{\bf Acknowledgements}

{\it We are deeply endebted to J. Polchinski for pointing out a major error
regarding the decoupling of BRST-trivial modes in string theory,
to T. Banks for elucidating a similar issue in the context
of mass renormalization, and to the above and to O. Aharony and M. Dine for
useful discussions on the framework.}
We thank many people for discussions, particularly O. Aharony,
N. Arkani-Hamed, T. Banks, M. Berkooz, J. Cline, S. Dimopoulos, J. Distler,
M. Fabinger, S. Giddings, J. Gomis,
A. Hashimoto, S. Hellerman, N. Kaloper, S. Kachru, D. Kaplan, M. Kleban,
E. Martinec, M. Peskin, J. Polchinski, S. Shenker, M. Strassler, L.
Susskind, W. Taylor, S. Thomas, N. Weiner, and B. Zwiebach.
We are grateful to O. Aharony, T. Banks, M. Berkooz,
M. Peskin, and J. Polchinski for comments on
the manuscript.  A.A. and E.S. are grateful to the Aspen Center for
Physics and its workshop organizers for hospitality
during completion of this work.
The research of A.A. and E.S. was supported by the DOE
under contract DE-AC03-76SF00515 and the A.P. Sloan
Foundation.  A.A. was supported also by an NSF Graduate
Fellowship.  The work of JM was supported in part by
National Science Foundation grant PHY-0097915, and by
a Princeton University Dicke Fellowship.

\listrefs

\end